\documentclass[%
reprint,
superscriptaddress,
amsmath,amssymb,
aps,
]{revtex4-2}
\usepackage[dvipdfmx]{graphicx}
\usepackage{dcolumn}
\usepackage{bm}
\usepackage{siunitx}
\usepackage{color}
\usepackage{braket}

\begin{document}
	
	\preprint{APS/123-QED}
	
	\title{Discontinuous Strong-to-Weak Symmetry Breaking Transition from \\Thermal Pure States}
	
	\author{Taiki Haga}
	\email{taiki.haga@omu.ac.jp}
	\affiliation{Department of Physics and Electronics, Osaka Metropolitan University, Sakai-shi, Osaka 599-8531, Japan}
	\author{Masaya Kunimi}
	\affiliation{Department of Physics, Tokyo University of Science, 1-3 Kagurazaka, Tokyo 162-8601, Japan}
	\author{Masaya Nakagawa}
	\affiliation{Department of Physics, The University of Tokyo, 7-3-1 Hongo, Tokyo 113-0033, Japan}
	
	\begin{abstract}
		We investigate the nonequilibrium dynamics of strong-to-weak spontaneous symmetry breaking in many-body quantum systems undergoing decoherence from thermal pure states.
		For generic initial pure states with volume-law entanglement entropy, we show that the system undergoes a discontinuous dynamical phase transition at a critical time.
		This transition is accompanied by a singularity in the entropy of the system, which saturates to its maximum value at the same critical time.
		Through numerical simulations of the dephasing Ising and hard-core boson models, we establish the universality of this transition across different symmetries.
		Our results reveal that the dynamical emergence of a decohered mixed state from a highly entangled state is not a gradual asymptotic relaxation, but rather a sharp phase transition driven by a sudden collapse of global coherence.
	\end{abstract}
	
	\maketitle
	
	{\em Introduction.--}
	The characterization of quantum phases of matter has recently been extended beyond pure states to encompass the rich structure of mixed quantum states in open systems. 
	A central concept in this frontier is strong-to-weak spontaneous symmetry breaking (SWSSB) \cite{Lee-23, Sala-24, Lessa-25, ChongWang-26, Kuno-24, Guo-25, Ellison-25, Luo-25, Ma-25, Shah-25, Orito-25, Zhang-25, Huang-25, Sa-25, Gu-25, Weinstein-25, Sun-25, Feng-25, Liu-25, Ando-26, Hauser-26, Shu-26, Wang-26, Teh-25, Ding-26}.
	In mixed states, symmetries manifest themselves in two distinct ways: \emph{strong symmetry}, where the symmetry is satisfied by each individual pure state in the ensemble, and \emph{weak symmetry}, which is only preserved by the ensemble as a whole \cite{Buca-12, Albert-14}.
	SWSSB occurs when a strong symmetry is spontaneously broken while the weak symmetry remains intact. 
	Unlike conventional symmetry breaking, which is diagnosed by standard two-point correlation functions, SWSSB is characterized by long-range order in observables that are nonlinear functions of the density matrix, such as the fidelity correlator and the R\'enyi-2 correlator \cite{Lee-23, Sala-24, Lessa-25, ChongWang-26, Weinstein-25, Liu-25}.
	
	While the emergence of SWSSB under decoherence from trivial product states and, more generally, states exhibiting area-law entanglement has been extensively investigated \cite{Lessa-25, Guo-25, Orito-25, Hauser-26, Shu-26}, it remains unclear how this symmetry-breaking order emerges when the starting point is a highly entangled pure state, such as a volume-law state generated by unitary dynamics.
	Understanding the degradation of many-body entanglement is not merely important for the advancement of quantum technologies but is deeply connected to a fundamental question in nonequilibrium statistical mechanics: Does the collapse of global coherence and the subsequent onset of SWSSB proceed via a gradual asymptotic decay, or does it show a sharp dynamical singularity in the thermodynamic limit?
	
	\begin{figure}[b]
		\centering
		\includegraphics[width=0.45\textwidth]{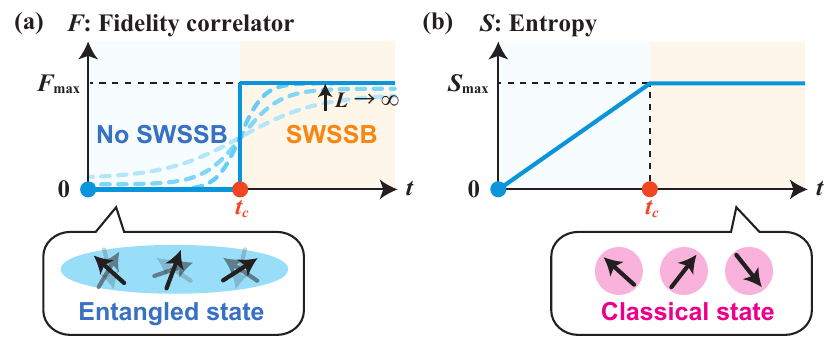}
		\caption{Schematic illustration of the dynamical emergence of SWSSB.
			(a) Time evolution of the fidelity correlator $F$. 
			In the thermodynamic limit ($L \to \infty$, solid line), the system undergoes a discontinuous dynamical phase transition at a finite critical time $t_c$, abruptly entering the SWSSB phase.
			(b) The corresponding time evolution of the global entropy $S$. 
			Starting from a highly entangled pure state, the entropy grows and exhibits a sharp kink at $t_c$, saturating to its maximum value $S_{\text{max}}$.}
		\label{fig_schematic}
	\end{figure}
	
	In this Letter, we investigate the nonequilibrium dynamics of SWSSB in highly entangled states under dephasing. 
	Through theoretical analysis and numerical simulations, we demonstrate that for generic initial states with volume-law entanglement \cite{Page-93}, the fidelity correlator and the R\'enyi-2 correlator exhibit a discontinuous jump from zero to a finite value at a critical time $t_c$ in the thermodynamic limit, signaling a first-order dynamical phase transition into the SWSSB phase [see Fig.~\ref{fig_schematic}(a)].
	Notably, this transition is accompanied by a singularity of the global entropy due to saturation to its maximum value [see Fig.~\ref{fig_schematic}(b)].
	Our result suggests that the choice of initial states is crucial for the critical dynamics of SWSSB in open quantum systems.

	{\em Strong-to-weak spontaneous symmetry breaking.--}
	In open quantum systems, symmetry of a mixed state $\rho$ is classified into two distinct forms \cite{Buca-12, Albert-14}.
	Consider a symmetry group $G$ and its unitary representation $U_g$ acting on the system's Hilbert space.
	A density matrix $\rho$ is said to possess a strong symmetry under $G$ if it satisfies $U_g \rho = e^{i\theta_g} \rho \: (\theta_g \in \mathbb{R})$ for all $g \in G$.
	Physically, this implies that every member of the ensemble $\rho$ carries the same symmetry charge.
	In contrast, a weak symmetry is defined by the condition $U_g \rho U_g^\dagger = \rho$ for all $g \in G$.
	This corresponds to an ensemble where individual states may carry different charges, but the total distribution remains invariant under the symmetry transformation.
	Note that strong symmetry leads to weak symmetry.
	SWSSB is defined as the phenomenon where a strong symmetry is spontaneously broken into a weak one, a transition that is unique to mixed quantum states.
	
	To detect SWSSB, we introduce nonlinear observables of the density matrix. 
	The most fundamental diagnostic is the fidelity correlator \cite{Lessa-25, ChongWang-26}.
	For a local operator $O_x$ acting on site $x$ and charged under the strong symmetry, the fidelity correlator is defined as $F_O(x,y) = F(\rho, O_x O_y^\dagger \rho O_y O_x^\dagger)$, where $F(\rho, \sigma) = \text{Tr}\sqrt{\sqrt{\rho}\sigma\sqrt{\rho}}$ is the quantum fidelity. 
	In the thermodynamic limit, SWSSB is characterized by the long-range order in the fidelity correlator, $\lim_{|x-y|\to\infty} F_O(x,y) > 0$, while the linear correlation function $\text{Tr}(\rho O_x O_y^\dagger)$ vanishes.
	Alternatively, one can employ the R\'enyi-2 correlator \cite{Lee-23, Sala-24, Lessa-25, ChongWang-26}, which is defined through the Choi-Jamio\l kowski isomorphism in a doubled Hilbert space: $R^{(2)}_O(x,y) = \text{Tr}(\rho O_x O_y^\dagger \rho O_y O_x^\dagger)/\text{Tr}(\rho^2)$.
	This corresponds to the correlation function of the operator $O_x \otimes O_x^*$ in the doubled space. 
	While $R^{(2)}_O(x,y)$ is more accessible for numerical and experimental evaluation, it is not always equivalent to the fidelity correlator \cite{Lessa-25}.
	
	{\em Discontinuous emergence of SWSSB.--}
	We consider the time evolution of a quantum many-body system governed by a Lindblad master equation \cite{Breuer, Rivas}:
	\begin{equation}
		\frac{d\rho}{dt} = -i[H, \rho] + \gamma \sum_n \left( L_n \rho L_n^\dagger - \frac{1}{2} \{L_n^\dagger L_n, \rho\} \right),
	\end{equation}
	where the Hamiltonian $H$ and the local Lindblad operators $L_n$ are assumed to be invariant under a symmetry group $G$, such that $[H, U_g] = 0$ and $[L_n, U_g] = [L_n^\dag, U_g] = 0$ for all $g \in G$.
	The parameter $\gamma$ denotes the strength of the coupling between the system and an environment.
	Under these conditions, if the initial state $\rho_0$ possesses a strong symmetry ($U_g \rho_0 = e^{i\theta_g} \rho_0$), this symmetry is preserved throughout the time evolution, i.e., $U_g \rho(t) = e^{i\theta_g} \rho(t)$ for finite-size systems and any $t > 0$.
	Furthermore, we assume that the dissipators lead the system toward an infinite-temperature steady state within the corresponding symmetry sector.
	Therefore, in the long-time limit, the system is guaranteed to exhibit a non-vanishing fidelity correlator.
	
	We prepare the system in a pure state $\rho_0 = |\psi_0\rangle\langle\psi_0|$ with the strong symmetry.
	For the purpose of our study, we assume that the initial state satisfies the following three properties:
	\begin{enumerate}
		\item \textit{Volume-law entanglement:} The state $|\psi_0\rangle$ possesses entanglement entropy that scales linearly with the volume of the subsystem.
		\item \textit{Absence of SSB:} The initial state does not exhibit any conventional long-range order.
		\item \textit{Random or thermal nature:} The state $|\psi_0\rangle$ is a typical state within the symmetric Hilbert space, such as a Haar-random pure state or a thermal pure state \cite{Sugiura-13, Mori-18} generated by chaotic unitary dynamics.
	\end{enumerate}
	The three conditions set a ``purely quantum" and disordered starting point, providing a rigorous baseline to observe the dynamical onset of SWSSB as the system relaxes toward the maximally mixed state.
	The central question we address below is how the SWSSB order emerges in the thermodynamic limit $L \to \infty$.
	
	We propose the following conjecture: For initial states satisfying the above conditions, the emergence of the SWSSB order is not gradual. 
	Instead, the fidelity correlator undergoes a discontinuous dynamical transition at a critical time $t_c$: 
	\begin{equation}
		\lim_{|x-y| \to \infty} \lim_{L \to \infty} F_O(x, y; t) = 
		\begin{cases}
			0 & (t < t_c), \\
			c > 0 & (t > t_c),
		\end{cases}
		\label{conjecture}
	\end{equation}
	where the constant $c$ is the value of the fidelity correlator for the steady state [see Fig.~\ref{fig_schematic}(a)].
	
	It is crucial to distinguish the dynamical transition discussed in this work from the SWSSB transitions previously reported in the literature \cite{Lessa-25, Hauser-26, Shu-26}.
	Previous studies have primarily focused on the emergence of SWSSB when starting from trivial product states subjected to local decoherence.
	In contrast, our work investigates the dynamics starting from highly entangled, strongly quantum initial states.
	This distinction leads to fundamentally different physical consequences.
	For instance, the previously studied SWSSB transitions are known to be absent in one-dimensional systems \cite{Hauser-26}.
	As discussed later, the dynamical phase transition we report occurs at a finite time even in one-dimensional settings.
	
	Our random pure initial states are locally indistinguishable from the infinite-temperature state, implying the absence of diffusion of local charges toward the steady state.
	Such initial states exhibit ``local SWSSB" \cite{Divi-26, Liu-26, Zhang-26, Tang-26} while lacking global SWSSB order.
	Thus, the dynamical transition at $t = t_c$ cannot be detected by any local probes.
	We note that this transition can be experimentally detected in ultracold atom platforms through the measurement of purity \cite{Kaufman-16}.
	
	{\em Model.--}
	Specifically, we consider a one-dimensional transverse-field Ising model of length $L$ with periodic boundary conditions, subject to local dephasing.
	The Hamiltonian is given by
	\begin{equation}
		H = - J \sum_{j=1}^L \sigma_j^z \sigma_{j+1}^z - h \sum_{j=1}^L \sigma_j^x,
	\end{equation}
	where $\sigma_j^\mu$ ($\mu = x, y, z$) are the Pauli matrices acting on site $j$, $J$ is the strength of the nearest-neighbor interaction, and $h$ denotes the transverse field strength.
	The dissipation due to the coupling with the environment is modeled by local dephasing Lindblad operators $L_j = \sigma_j^x$ acting on every site $j=1, \dots, L$.
	This system is invariant under the global spin-flip unitary operator $X = \prod_{j=1}^L \sigma_j^x$, satisfying the commutation relations $[H, X] = 0$ and $[L_j, X] = 0$.
	Consequently, the dynamics preserve a strong $\mathbb{Z}_2$ symmetry.
	Starting from a strongly symmetric initial state ($X \rho_0 = \rho_0$), the dephasing drives the system toward the maximally mixed state $\rho(\infty)=(I + X) / 2^L$ within the symmetry sector.
	To probe the SWSSB order, we consider the R\'enyi-2 and fidelity correlators for $O_j = \sigma_j^z$.
	Furthermore, to demonstrate the universality of the dynamical phase transition, we provide additional results in the Supplemental Material \cite{SM} for a one-dimensional hard-core boson model under dephasing \cite{Pichler-10, Pichler-13, Sarkar-14, Luschen-17, Bouganne-20}, which possesses a strong $U(1)$ symmetry.

	{\em Theoretical analysis.--}
	We first present a theoretical analysis to show the discontinuous transition to the SWSSB phase from typical pure states.
	For analytical tractability, we focus on the R\'enyi-2 correlator.
	The system is initially prepared in a strongly symmetric random pure state. 
	Specifically, the initial state $\ket{\psi_0}$ is constructed by sampling a Haar-random pure state $\ket{\psi_{\text{Haar}}}$ from the entire Hilbert space and subsequently projecting it onto the even-parity sector: $\ket{\psi_0} \propto P_+ \ket{\psi_{\text{Haar}}}$, where $P_+ = (I + X)/2$.
	
	Note that the fluctuations of purity $\mathrm{Tr}[\rho(t)^2]$ over different realizations of the initial state are exponentially small compared to its ensemble average \cite{SM, Bao-26}.
	This self-averaging property allows us to express the typical value of the R\'enyi-2 correlator as the ratio of the ensemble averages:
	\begin{equation}
		R_O^{(2)}(i, j; t) = \frac{\mathbb{E}_{\psi_0}[\mathrm{Tr}[\rho(t) \sigma_{i}^z \sigma_{j}^z \rho(t) \sigma_{i}^z \sigma_{j}^z]]}{\mathbb{E}_{\psi_0}[\mathrm{Tr}[\rho(t)^2]]},
		\label{renyi_2_Ising}
	\end{equation}
	where $\mathbb{E}_{\psi_0}[\cdots]$ denotes the ensemble average over the initial states.
	To evaluate the moments of the density matrix, we employ the doubled Hilbert space $\mathcal{H} \otimes \mathcal{H}$ and introduce the swap operator $\mathcal{S}$, defined by $\mathcal{S}(\ket{u}\otimes\ket{v})=\ket{v}\otimes\ket{u}$ for any vectors $\ket{u}, \ket{v} \in \mathcal{H}$ \cite{Collins-10, Hamma-12, Roberts-17, Nahum-18}.
	Let $\Lambda_t$ be the quantum channel generated by the Lindblad master equation such that $\rho(t)=\Lambda_t[\rho_0]$.
	
	A key observation is that for $i \neq j$, the correlator in Eq.~\eqref{renyi_2_Ising} can be rewritten as
	\begin{equation}
		R_O^{(2)}(i, j; t) = \frac{2^{L-1}}{2^{L-1} + \mathrm{Tr}_{12}[\mathcal{S} (\Lambda_t \otimes \Lambda_t)[\mathcal{S}_+]]},
		\label{renyi_2_Ising_off_diagonal_1}
	\end{equation}
	where $\mathcal{S}_+=\mathcal{S} (P_+ \otimes P_+)$ is the swap operator restricted to the even-parity sector and $\mathrm{Tr}_{12}$ represents the trace over the doubled space $\mathcal{H} \otimes \mathcal{H}$ (see the End Matter for a detailed derivation).
	We assume that there exists a function $f(t)$ such that $\lim_{L \to \infty} \mathrm{Tr}_{12}[\mathcal{S} (\Lambda_t \otimes \Lambda_t)[\mathcal{S}_+]] / f(t)^{L-1} = 1$.
	The function $f(t)$ captures the typical decay dynamics of purity under dephasing.
	
	Since $\mathrm{Tr}_{12}[\mathcal{S} (\Lambda_t \otimes \Lambda_t)[\mathcal{S}_+]]$ evolves from an initial value of $4^{L-1}$ to a steady-state value of $1$ as $t \to \infty$, it follows that $f(0)=4$ and $f(\infty)=1$.
	In the thermodynamic limit ($L \to \infty$), Eq.~\eqref{renyi_2_Ising_off_diagonal_1} becomes
	\begin{equation}
		R_O^{(2)}(i, j; t) = \frac{1}{1 + [f(t)/2]^{L-1}} \to 
		\begin{cases}
			0 & (f(t) > 2), \\
			1 & (f(t) < 2).
		\end{cases}
		\label{renyi_2_Ising_off_diagonal_2}
	\end{equation}
	Therefore, provided that $f(t)$ crosses the threshold value of $2$ monotonically, $R_O^{(2)}(i, j; t)$ must exhibit a discontinuous jump from zero to unity at the critical time $t_c$ satisfying $f(t_c)=2$.
	An explicit calculation of $R_O^{(2)}(i, j; t)$ using a cluster mean-field approximation is provided in the Supplemental Material \cite{SM}.
	
	\begin{figure}
		\centering
		\includegraphics[width=0.45\textwidth]{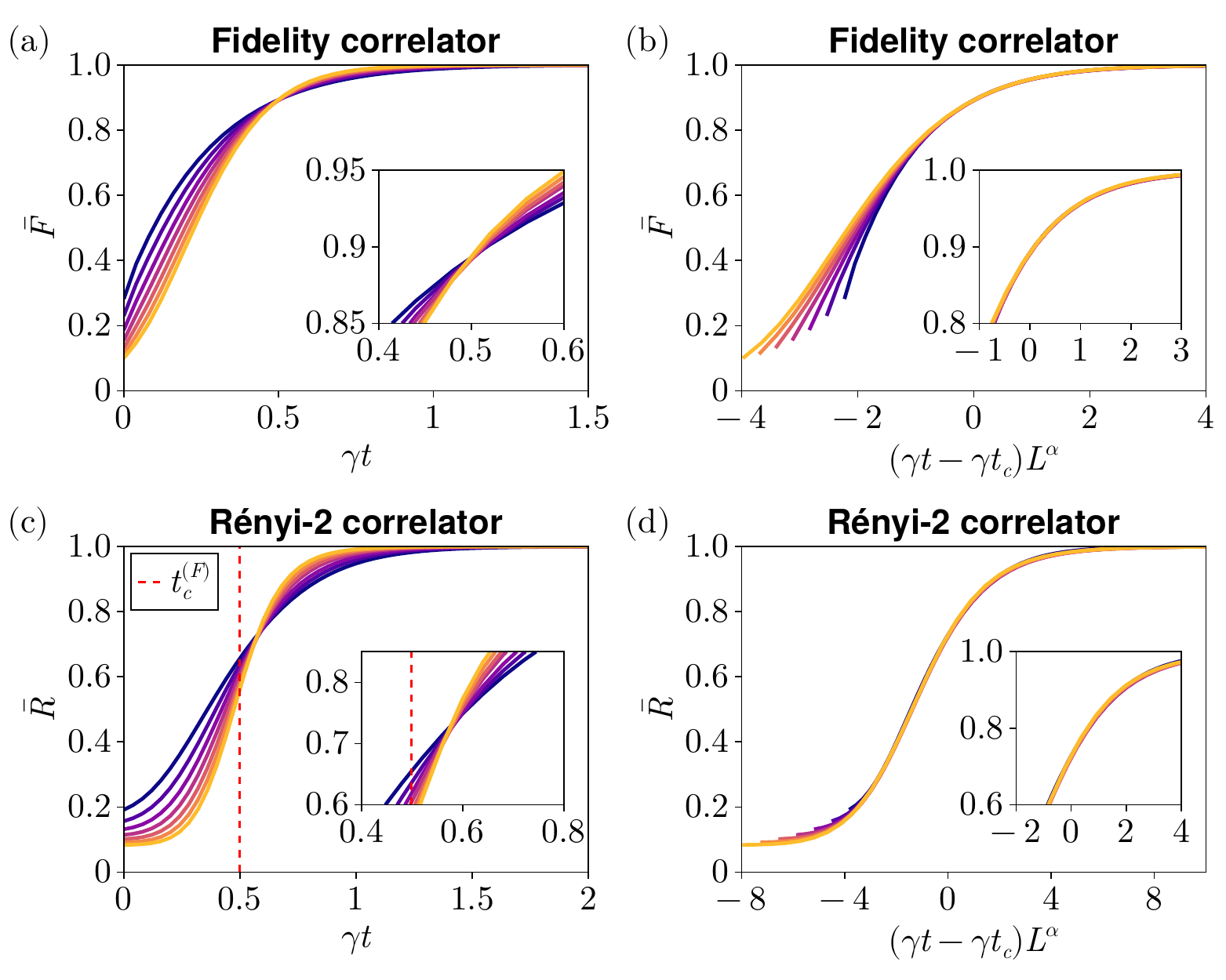}
		\caption{Dynamical phase transitions of the fidelity correlator and the R\'enyi-2 correlator.
			(a) Time evolution of the spatially averaged fidelity correlator $\bar{F}$ as a function of dimensionless time $\gamma t$.
			The parameters are set to $J/\gamma=1$ and $h=0$.
			Colors from dark to light represent system sizes $L = 6, 7, 8, 9, 10, 11,$ and $12$. 
			The inset highlights the crossing point of the trajectories for different $L$, identifying the critical time $t_c^{(F)}$ of the dynamical transition.
			(b) Finite-size scaling analysis of the data in (a). 
			All curves collapse onto a single universal scaling function when plotted against the rescaled time $(\gamma t - \gamma t_c)L^\alpha$ with $\alpha = 0.84$.
			(c) Time evolution of the spatially averaged R\'enyi-2 correlator $\bar{R}$.
			The vertical red dashed line indicates the critical time $t_c^{(F)}$ extracted from the fidelity correlator.
			The inset highlights the crossing point of $\bar{R}$, identifying the critical time $t_c^{(R)}$, which occurs later than $t_c^{(F)}$.
			(d) Finite-size scaling analysis of the data in (c) using a scaling exponent $\alpha = 1.05$.}
		\label{fig_fidelity_renyi_correlator}
	\end{figure}
	
	We next discuss the relationship between the R\'enyi-2 correlator and the global R\'enyi-2 entropy $S_2(t)=-\ln \mathrm{Tr}[\rho(t)^2]$.
	Due to the self-averaging property \cite{SM}, the typical value of the entropy can be expressed as $S_2(t) \simeq -\ln \mathbb{E}_{\psi_0}[\mathrm{Tr}[\rho(t)^2]]$.
	The time evolution of the purity is given by
	\begin{equation}
		\mathbb{E}_{\psi_0}[\mathrm{Tr}[\rho(t)^2]] = \frac{2^{L-1} + f(t)^{L-1}}{2^{L-1} (2^{L-1} + 1)}.
		\label{purity_evolution}
	\end{equation}
	For $t < t_c$, the second term in the numerator dominates in the limit $L \to \infty$ because $f(t) > 2$, leading to $\mathbb{E}_{\psi_0}[\text{Tr}[\rho(t)^2]] \simeq (f(t)/4)^{L-1}$.
	The normalized R\'enyi-2 entropy $\tilde{S}_2(t) = S_2(t) / [(L-1) \ln 2]$ thus behaves as $\tilde{S}_2(t) \simeq 2 - \ln f(t) / \ln 2$.
	Conversely, for $t > t_c$, the first term dominates because $f(t) < 2$, yielding $\mathbb{E}_{\psi_0}[\text{Tr}[\rho(t)^2]] \simeq 2^{-(L-1)}$ and thus $\tilde{S}_2(t) \simeq 1$.
	This argument implies that the R\'enyi-2 entropy saturates to its maximum value precisely at the dynamical transition point of the corresponding correlator [see Fig.~\ref{fig_schematic}(b)].

	{\em Numerical results.--}
	We present numerical results obtained through exact integration of the Lindblad master equation.
	To quantify the emergence of SWSSB, we track the spatially averaged fidelity correlator and R\'enyi-2 correlator, defined as $\bar{F} = L^{-2} \sum_{i,j} F_O(i, j)$ and $\bar{R} = L^{-2} \sum_{i,j} R_O^{(2)}(i, j)$.
	
	Figure \ref{fig_fidelity_renyi_correlator}(a) shows the time evolution of $\bar{F}$ for different system sizes $L$.
	The value of $\bar{F}$ remains near zero at early times and grows toward unity as decoherence proceeds. 
	We observe a clear crossing of the trajectories of $\bar{F}$ for different $L$ at a critical time $t_c^{(F)}$, estimated as $\gamma t_c^{\text{(F)}} \simeq 0.50$.
	To elucidate the nature of this transition, we perform a finite-size scaling analysis using the ansatz $\bar{F} = \Phi((t-t_c)L^\alpha)$.
	As shown in Fig.~\ref{fig_fidelity_renyi_correlator}(b), the data for all system sizes collapse onto a single universal curve $\Phi(x)$, which monotonically grows from zero to unity as $x$ increases from $-\infty$ to $\infty$.
	This robust collapse supports that the onset of SWSSB is a discontinuous dynamical phase transition, where the fidelity correlator jumps from zero to a finite value at $t_c^{(F)}$ in the thermodynamic limit.
	
	A similar dynamical transition is observed for the spatially averaged R\'enyi-2 correlator $\bar{R}$, as shown in Fig.~\ref{fig_fidelity_renyi_correlator}(c).
	We find that $\bar{R}$ also exhibits a crossing at a critical time $t_c^{(R)}$, where $\gamma t_c^{(R)} \simeq 0.58$.
	Notably, the critical time $t_c^{(F)}$ extracted from the fidelity correlator precedes that of the R\'enyi-2 correlator ($t_c^{(F)} < t_c^{(R)}$).
	This tendency of the R\'enyi-2 quantities to underestimate the SWSSB phase boundary is consistent with previous observations in the toric code under bit-flip noise \cite{Lessa-25}.
	As detailed in the Supplemental Material \cite{SM}, qualitatively the same behavior is observed for the hard-core boson model.
	
	\begin{figure}
		\centering
		\includegraphics[width=0.45\textwidth]{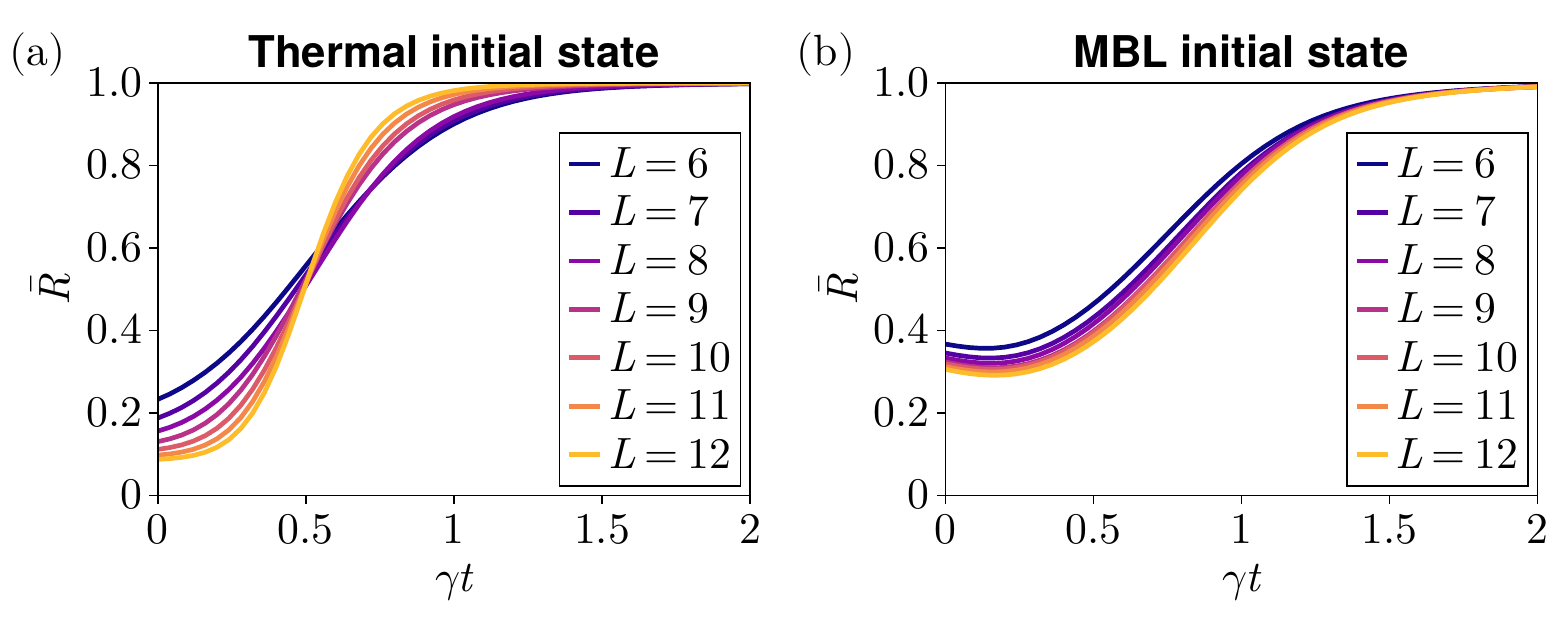}
		\caption{Time evolution of the spatially averaged R\'enyi-2 correlator $\bar{R}$ starting from different initial eigenstates of a random-field Ising model.
			(a) For a thermal initial state (weak disorder), the curves for different system sizes $L$ exhibit a crossing point, indicating a discontinuous dynamical transition.
			(b) For a many-body localized (MBL) initial state (strong disorder), the crossing point is absent.}
		\label{fig_renyi_2_correlator_thermal_MBL_initial_state}
	\end{figure}
	
	To highlight the crucial role of the initial entanglement structure, we investigate the dynamics starting from different types of excited eigenstates.
	We prepare an initial state $|\psi_0\rangle$ by exact diagonalization of an auxiliary non-integrable Ising Hamiltonian with random magnetic fields \cite{SM}.
	By tuning the disorder strength, we can generate either a thermal pure state satisfying the eigenstate thermalization hypothesis (exhibiting volume-law entanglement) \cite{Deutsch-91, Srednicki-94, Rigol-08} or a many-body localized (MBL) state (exhibiting area-law entanglement) \cite{Nandkishore-15, Abanin-19} from the middle of the spectrum.
	
	Figure \ref{fig_renyi_2_correlator_thermal_MBL_initial_state} illustrates the time evolution of $\bar{R}$ for these two contrasting initial conditions. 
	When the initial state is a thermal pure state [Fig.~\ref{fig_renyi_2_correlator_thermal_MBL_initial_state}(a)], $\bar{R}$ exhibits a crossing point for different system sizes, consistent with the discontinuous dynamical transition observed for Haar-random states.
	In contrast, when the system is initialized in an MBL state [Fig.~\ref{fig_renyi_2_correlator_thermal_MBL_initial_state}(b)], the crossing behavior disappears.
	These results suggest that a volume-law entangled initial state is a necessary condition for the sharp onset of SWSSB.
	
	\begin{figure}
		\centering
		\includegraphics[width=0.45\textwidth]{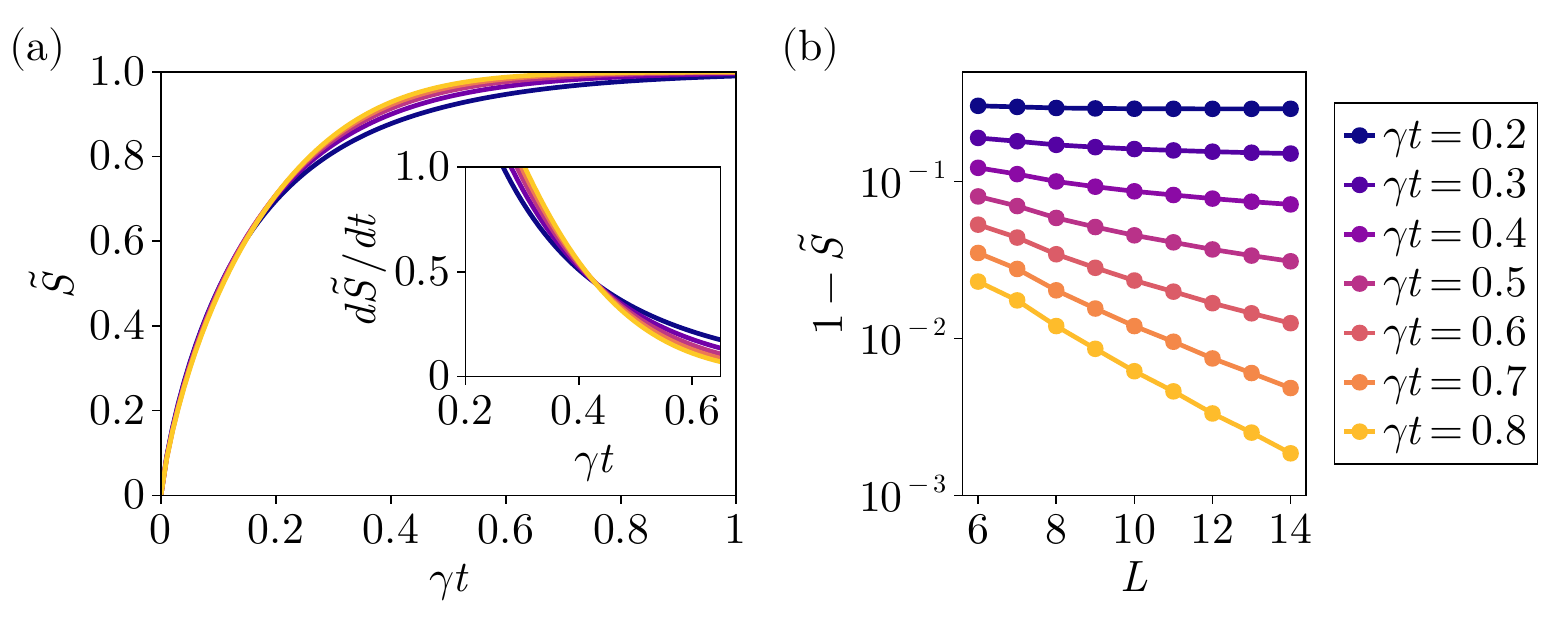}
		\caption{Saturation of the global entropy.
			(a) Time evolution of the normalized global entropy $\tilde{S}$.
			Colors from dark to light represent system sizes $L = 6, 8, 10, 12,$ and $14$.
			The inset shows the time derivative $d\tilde{S}/dt$, where the curves for different system sizes cross at a critical time $t_{\text{sat}}$.
			(b) System-size dependence of $1-\tilde{S}$.
			For $t>t_{\text{sat}}$, the value exhibits an exponential decay, $1-\tilde{S} \sim e^{-O(L)}$.}
		\label{fig_entropy_random_initial_state}
	\end{figure}
	
	In the theoretical analysis, we have shown that the SWSSB transition in the R\'enyi-2 correlator is accompanied by the saturation of the global R\'enyi-2 entropy.
	Here, we extend our analysis to the global von Neumann entropy $S(t) = -\text{Tr}[\rho(t) \ln \rho(t)]$.
	As shown in Fig.~\ref{fig_entropy_random_initial_state}(a), the normalized entropy $\tilde{S}(t) = S(t)/[(L-1)\ln 2]$ monotonically increases toward unity.
	Crucially, the time derivative $d\tilde{S}/dt$ [inset of Fig.~\ref{fig_entropy_random_initial_state}(a)] shows a crossing point for different system sizes, pinpointing a critical time $t_{\text{sat}}$ at which $d\tilde{S}/dt$ exhibits a discontinuous jump in the thermodynamic limit.
	Figure \ref{fig_entropy_random_initial_state}(b) shows the system-size dependence of the entropy deficit, $1-\tilde{S}(t)$. 
	While $1-\tilde{S}(t)$ converges to a nonzero constant for $t < t_{\text{sat}}$, it decays exponentially, $1-\tilde{S}(t) \sim e^{-O(L)}$, for $t > t_{\text{sat}}$.
	These results suggest that the global quantum coherence of the initial state is completely destroyed after $t_{\text{sat}}$.
	
	While it is tempting to conjecture that the saturation time $t_{\text{sat}}$ of the von Neumann entropy coincides with the critical time $t_c^{(F)}$ of the fidelity correlator, the numerically determined $\gamma t_c^{(F)} \simeq 0.50$ is slightly larger than the observed $\gamma t_{\text{sat}} \simeq 0.43$ [see the inset of Fig.~\ref{fig_entropy_random_initial_state}(a)].
	This discrepancy may stem from finite-size effects or reflect a fundamental difference between their nonlinear structures ($\ln \rho$ versus $\sqrt{\rho}$).

	{\em Discussion.--}
	To clarify the necessary conditions for this dynamical transition, we investigate the dynamics starting from different types of initial states.
	When the system is initialized in an area-law entangled state, such as the $x$-polarized GHZ state, or a log-law entangled state, such as the Dicke state \cite{Popkov-05, Kunimi-25}, the SWSSB order parameter does not exhibit a discontinuous jump.
	Instead, the onset time $t_c$ at which the order parameter begins to grow diverges logarithmically with the system size $L$ \cite{SM}, being consistent with the behavior reported in Ref.~\cite{Shu-26}.
	Consequently, in the thermodynamic limit, the order parameter remains pinned at zero, indicating the absence of a finite-time transition.
	Notably, we observe qualitatively identical behavior when the initial state is a rainbow state \cite{SM}, which possesses volume-law entanglement but is non-thermal \cite{Langlett-22, Wildeboer-22, Chiba-24, Mohapatra-25}.
	These findings demonstrate that both volume-law entanglement and the thermal nature are indispensable prerequisites for the finite-time dynamical transition into the SWSSB phase.
	
	We remark on the fundamental difference in the detection capabilities of the two metrics. 
	Consider a mixed initial state composed of a Haar-random state and a GHZ state: $\rho_0 = (\rho_{\text{rand}}+\rho_{\text{GHZ}})/2$.
	In this scenario, the R\'enyi-2 correlator exhibits behavior almost identical to that of the pure GHZ case.
	To see this, note that, under decoherence, the purity of the random state decays exponentially, whereas that of the GHZ state remains $O(1)$ ($\text{Tr}[\rho_{\text{GHZ}}(t)^2] \gg \text{Tr}[\rho_{\text{rand}}(t)^2]$).
	Consequently, the R\'enyi-2 correlator is dominated by the GHZ component and remains pinned at zero in the thermodynamic limit.
	In contrast, the joint concavity of the fidelity leads to $\bar{F}(\rho(t)) \ge \bar{F}(\rho_{\text{rand}}(t))/2 + \bar{F}(\rho_{\text{GHZ}}(t))/2$.
	For $t > t_c$ in the thermodynamic limit, while the GHZ component $\bar{F}(\rho_{\text{GHZ}}(t))$ remains zero, the random component $\bar{F}(\rho_{\text{rand}}(t))$ exhibits a discontinuous jump to a finite value.
	Thus, we have $\bar{F}(\rho(t)) > 0$ for $t > t_c$.
	The fact that the fidelity correlator can remain finite even when the R\'enyi-2 correlator vanishes explains why $t_c$ detected by the fidelity correlator can precede that of the R\'enyi-2 correlator for random initial states.
	
	It is worth noting that Refs.~\cite{Chen-25, Chen-20, Wang-24} investigated similar SWSSB and entanglement transitions in the dissipative Sachdev-Ye-Kitaev model.
	Our work fundamentally differs from the previous works in two aspects.
	First, rather than relying on all-to-all random interactions, we demonstrate this dynamical transition in generic finite-dimensional systems with local interactions.
	Second, instead of employing initial states entangled with external auxiliary systems \cite{Chen-25, Chen-20, Wang-24}, we focus on the degradation of genuine internal spatial entanglement.
	These distinctions establish that the discontinuous onset of SWSSB is not unique to infinite-range models, but a universal information-theoretic singularity governed by the intrinsic structure of many-body entanglement.

	{\em Acknowledgments.--}
	This work was supported by JSPS KAKENHI Grants No.~JP22K13983 (T.H.), No.~JP25K00215 (M.K.), JST ASPIRE No.~JPMJAP24C2 (M.K.), and No.~JP24K16989 (M.N.).
	
	{\em Data availability.--}
	The data and codes that support the findings of this paper are openly available \cite{data}.

	\section*{End Matter}
	
	In this End Matter, we provide a detailed derivation of Eq.~\eqref{renyi_2_Ising_off_diagonal_1} for the R\'enyi-2 correlator of the dephasing Ising model.
	The objective is to evaluate the following quantity:
	\begin{equation}
		R_O^{(2)}(i, j; t) = \frac{\mathbb{E}_{\psi_0}[\mathrm{Tr}[\rho(t) \sigma_{i}^z \sigma_{j}^z \rho(t) \sigma_{i}^z \sigma_{j}^z]]}{\mathbb{E}_{\psi_0}[\mathrm{Tr}[\rho(t)^2]]}.
		\label{appendix_renyi_2_Ising}
	\end{equation}
	To treat the products of density matrices, we utilize the doubled Hilbert space $\mathcal{H} \otimes \mathcal{H}$ and the swap operator $\mathcal{S}$, defined by $\mathcal{S}(\ket{u}\otimes\ket{v})=\ket{v}\otimes\ket{u}$. 
	For any operators $A$ and $B$, the following trace identity holds:
	\begin{equation}
		\mathrm{Tr}_{12}[(A \otimes B) \mathcal{S}] = \mathrm{Tr}[AB],
	\end{equation}
	where $\mathrm{Tr}_{12}$ represents the trace over the doubled space $\mathcal{H} \otimes \mathcal{H}$.
	Using this identity, the denominator and numerator of Eq.~\eqref{appendix_renyi_2_Ising} can be expressed as
	\begin{equation}
		\mathrm{Tr}[\rho(t)^2] = \mathrm{Tr}_{12}[(\rho(t) \otimes \rho(t)) \mathcal{S}],
		\label{appendix_denominator_1}
	\end{equation}
	\begin{equation}
		\begin{split}
			&\mathrm{Tr}[\rho(t) \sigma_i^z \sigma_j^z \rho(t) \sigma_i^z \sigma_j^z] \\
			&= \mathrm{Tr}_{12}[(\rho(t) \otimes \rho(t)) (\sigma_i^z \sigma_j^z \otimes \sigma_i^z \sigma_j^z) \mathcal{S}].
		\end{split}
		\label{appendix_numerator_1}
	\end{equation}
	Let $\Lambda_t$ be the quantum channel generated by the Lindblad master equation such that $\rho(t)=\Lambda_t[\rho_0]$.
	Then, Eqs.~\eqref{appendix_denominator_1} and \eqref{appendix_numerator_1} can be rewritten as
	\begin{equation}
		\mathrm{Tr}[\rho(t)^2] = \mathrm{Tr}_{12}[\mathcal{S} (\Lambda_t \otimes \Lambda_t)[\rho_0 \otimes \rho_0]],
		\label{appendix_denominator_2}
	\end{equation}
	\begin{equation}
		\begin{split}
			&\mathrm{Tr}[\rho(t) \sigma_i^z \sigma_j^z \rho(t) \sigma_i^z \sigma_j^z] \\
			&= \mathrm{Tr}_{12}[(\sigma_i^z \sigma_j^z \otimes \sigma_i^z \sigma_j^z) \mathcal{S} (\Lambda_t \otimes \Lambda_t)[\rho_0 \otimes \rho_0]].
		\end{split}
		\label{appendix_numerator_2}
	\end{equation}
	
	Let us evaluate the ensemble average over the initial pure state $\rho_0 = |\psi_0\rangle\langle\psi_0|$.
	The initial state $\ket{\psi_0}$ is constructed by sampling a Haar-random pure state $\ket{\psi_{\text{Haar}}}$ from the entire Hilbert space and subsequently projecting it onto the even-parity sector: $\ket{\psi_0} \propto P_+ \ket{\psi_{\text{Haar}}}$, where $P_+ = (I + X)/2$.
	The ensemble average of the doubled initial state is given by
	\begin{equation}
		\mathbb{E}_{\psi_0}[\rho_0 \otimes \rho_0] = \frac{P_+ \otimes P_+ + \mathcal{S}_+}{D_+ (D_+ + 1)},
	\end{equation}
	where $\mathcal{S}_+=\mathcal{S} (P_+ \otimes P_+)$ is the swap operator restricted to the even-parity sector and $D_+=2^{L-1}$ is its dimension \cite{Collins-10, Hamma-12, Roberts-17, Nahum-18}.
	Substituting this into the ensemble average of Eqs.~\eqref{appendix_denominator_2} and \eqref{appendix_numerator_2}, we obtain
	\begin{equation}
		\begin{split}
			&\mathbb{E}_{\psi_0}[\mathrm{Tr}[\rho(t)^2]] = \frac{1}{D_+ (D_+ + 1)} \\
			&\times \mathrm{Tr}_{12}[\mathcal{S} (\Lambda_t \otimes \Lambda_t)[P_+ \otimes P_+ + \mathcal{S}_+]],
		\end{split}
		\label{appendix_denominator_3}
	\end{equation}
	
	\begin{equation}
		\begin{split}
			&\mathbb{E}_{\psi_0}[\mathrm{Tr}[\rho(t) \sigma_i^z \sigma_j^z \rho(t) \sigma_i^z \sigma_j^z]] = \frac{1}{D_+ (D_+ + 1)} \\
			&\times \mathrm{Tr}_{12}[(\sigma_i^z \sigma_j^z \otimes \sigma_i^z \sigma_j^z) \mathcal{S} (\Lambda_t \otimes \Lambda_t)[P_+ \otimes P_+ + \mathcal{S}_+]].
		\end{split}
		\label{appendix_numerator_3}
	\end{equation}
	
	We define
	\begin{align}
		A(t)&:=\mathrm{Tr}_{12}[\mathcal{S} (\Lambda_t \otimes \Lambda_t)[\mathcal{S}_+]], \\
		B_{ij}(t)&:=\mathrm{Tr}_{12}[(\sigma_i^z \sigma_j^z \otimes \sigma_i^z \sigma_j^z) \mathcal{S} (\Lambda_t \otimes \Lambda_t)[\mathcal{S}_+]].
	\end{align}
	By noting that
	\begin{align}
		\mathrm{Tr}_{12}[\mathcal{S} (\Lambda_t \otimes \Lambda_t)[P_+ \otimes P_+]] = D_+, \\
		\mathrm{Tr}_{12}[(\sigma_i^z \sigma_j^z \otimes \sigma_i^z \sigma_j^z) \mathcal{S} (\Lambda_t \otimes \Lambda_t)[P_+ \otimes P_+]] = D_+,
	\end{align}
	the R\'enyi-2 correlator simplifies to
	\begin{equation}
		R_O^{(2)}(i, j; t) = \frac{D_+ + B_{ij}(t)}{D_+ + A(t)}.
	\end{equation}
	
	For $i \neq j$, $B_{ij}(t)$ remains negligible compared to $D_+$ throughout the dynamics.
	In fact, $B_{ij}(0)=0$ for the initial state.
	In the long-time limit, the swap operator evolves as
	\begin{equation}
		\lim_{t \to \infty} \: (\Lambda_t \otimes \Lambda_t)[\mathcal{S}_+] = \frac{P_+ \otimes P_+}{D_+}.
		\label{appendix_S_evolution_long_time_limit}
	\end{equation}
	Thus, it follows that $B_{ij}(\infty)=1$.
	Assuming $B_{ij}(t)/D_+=O(1/D_+)$ for all $t$, we can safely neglect $B_{ij}(t)$ in the thermodynamic limit, yielding
	\begin{equation}
		R_O^{(2)}(i, j; t) = \frac{D_+}{D_+ + A(t)},
	\end{equation}
	which is the expression used in the main text.
	Note that $A(0)=D_+^2$ and $A(\infty)=1$.
	Further details on the explicit calculation of $A(t)$ via a cluster mean-field approximation are provided in the Supplemental Material \cite{SM}.
	Finally, the expression for the purity in Eq.~\eqref{purity_evolution} can be obtained from Eq.~\eqref{appendix_denominator_3}.
	
	\clearpage
	\newpage
	
	\setcounter{equation}{0}
	\setcounter{figure}{0}
	\setcounter{table}{0}
	\setcounter{page}{1}
	\setcounter{section}{0}
	
	\renewcommand{\theequation}{S\arabic{equation}}
	\renewcommand{\thefigure}{S\arabic{figure}}
	\renewcommand{\thetable}{S\arabic{table}}
	\renewcommand{\thepage}{S\arabic{page}}
	\renewcommand{\thesection}{S\arabic{section}}
	
	\onecolumngrid
	\begin{center}
		\textbf{\large Supplemental Material for \\ ``Discontinuous Strong-to-Weak Symmetry Breaking Transition from \\Thermal Pure States"} \\
		\vspace{4mm}
		Taiki Haga,$^{1}$ Masaya Kunimi,$^{2}$ and Masaya Nakagawa$^{3}$ \\
		\vspace{2mm}
		\small
		$^{1}$\textit{Department of Physics and Electronics, Osaka Metropolitan University, Sakai-shi, Osaka 599-8531, Japan} \\
		$^{2}$\textit{Department of Physics, Tokyo University of Science, 1-3 Kagurazaka, Tokyo 162-8601, Japan} \\
		$^{3}$\textit{Department of Physics, The University of Tokyo, 7-3-1 Hongo, Tokyo 113-0033, Japan} \\
		\vspace{4mm}
	\end{center}
	
	This Supplemental Material provides technical details and additional theoretical analyses to support the findings in the main text.
	In Sec.~\ref{sec:cmf}, we present detailed calculations of the R\'enyi-2 correlator and the R\'enyi-2 entropy using a cluster mean-field approximation.
	In Sec.~\ref{sec:variance}, we provide analytical and numerical estimations of the fluctuations in purity to verify the self-averaging property of the dynamics.
	In Sec.~\ref{sec:num_details}, we describe the numerical methods employed to simulate the dissipative dynamics of the Ising model under dephasing.
	In Sec.~\ref{sec:hard_core_boson}, we present supplementary results for the hard-core boson model to demonstrate the universality of the dynamical transition across different symmetry classes.
	In Sec.~\ref{sec:nonthermal}, we provide numerical evidence that the sharp dynamical transition is absent for non-thermal initial states, such as area-law and subvolume-law entangled states.

	\section{Cluster Mean-Field Approximation for the R\'enyi-2 Correlator}
	\label{sec:cmf}
	
	In the main text, we have introduced the following quantity: 
	\begin{equation}
		A(t) := \text{Tr}_{12}[\mathcal{S} \: (\Lambda_t \otimes \Lambda_t)[\mathcal{S}_+]] = \text{Tr}_{12}[(\Lambda_t^\dag \otimes \Lambda_t^\dag)[\mathcal{S}] \: \mathcal{S}_+],
	\end{equation}
	where $\mathcal{S}_+=\mathcal{S} \: (P_+ \otimes P_+)$ is the swap operator restricted to the even-parity sector, $P_+ = (I + X)/2$ is the projection operator onto this sector with the global parity operator $X = \prod_{j=1}^L \sigma_j^x$, and $\Lambda_t$ is the time evolution map.
	The global swap operator $\mathcal{S}$ is expressed as the product of local swap operators, $\mathcal{S} = \bigotimes_{j=1}^L S_j$.
	The R\'enyi-2 correlator is then given by
	\begin{equation}
		R_O^{(2)}(i, j; t) = \frac{1}{1 + A(t)/2^{L-1}}, \quad (i \neq j),
		\label{renyi_correlator_off_diagonal}
	\end{equation}
	yielding the spatially averaged R\'enyi-2 correlator,
	\begin{equation}
		\bar{R}(t) = \frac{1}{L} + \frac{L-1}{L} \frac{1}{1 + A(t)/2^{L-1}}.
		\label{spatially_averaged_renyi_correlator}
	\end{equation}
	The discontinuous transition of $\bar{R}(t)$ occurs when $A(t)$ crosses the threshold at which $A(t)^{1/(L-1)}/2 = 1$.
	
	Let us define the time-evolved swap operator in the doubled space as 
	\begin{equation}
		\mathcal{S}(t) := (\Lambda_t^\dag \otimes \Lambda_t^\dag)[\mathcal{S}].
		\label{time_evolved_swap_operator}
	\end{equation}
	This time-evolved swap operator obeys the adjoint Lindblad master equation extended to the doubled Hilbert space:
	\begin{equation}
		\frac{d\mathcal{S}(t)}{dt} = i \sum_{\alpha=1,2} [H^{(\alpha)}, \mathcal{S}(t)] + \gamma \sum_{\alpha=1,2} \sum_{j=1}^L \left( \sigma_{j, \alpha}^x \mathcal{S}(t) \sigma_{j, \alpha}^x - \mathcal{S}(t) \right),
		\label{master_equation_swap_operator}
	\end{equation}
	where $H^{(\alpha)}$ and $\sigma_{j,\alpha}^\mu$ act on the $\alpha$th copy of the system ($\alpha=1,2$).
	By expanding $P_+ = (I + X)/2$, $A(t)$ is rewritten as
	\begin{equation}
		A(t) = \frac{1}{4} (A_{II}(t) + A_{XI}(t) + A_{IX}(t) + A_{XX}(t)),
	\end{equation}
	\begin{equation}
		A_{II}(t) := \text{Tr}_{12}[\mathcal{S}(t) \: \mathcal{S}],
	\end{equation}
	\begin{equation}
		A_{XI}(t) := \text{Tr}_{12}[\mathcal{S}(t) \: \mathcal{S} \: (X \otimes I)],
	\end{equation}
	\begin{equation}
		A_{IX}(t) := \text{Tr}_{12}[\mathcal{S}(t) \: \mathcal{S} \: (I \otimes X)],
	\end{equation}
	\begin{equation}
		A_{XX}(t) := \text{Tr}_{12}[\mathcal{S}(t) \: \mathcal{S} \: (X \otimes X)].
	\end{equation}
	
	While the existence of the strong-to-weak spontaneous symmetry breaking (SWSSB) transition can be established without assuming a specific functional form for $A(t)$, its explicit evaluation is instructive for understanding the purity decay under decoherence.
	In the following, we provide a calculation of $A(t)$ by employing a cluster mean-field approximation.
	
	\subsection{Single-site mean-field approximation}
	
	First, we illustrate why the single-site mean-field approximation fails to reproduce the discontinuous dynamical transition.
	Within the single-site mean-field approximation, we assume a fully factorized ansatz:
	\begin{equation}
		\mathcal{S}(t) \simeq \bigotimes_{j=1}^L \Phi_j(t),
		\label{single_site_mean_field_ansatz}
	\end{equation}
	where $\Phi_j(t)$ is an operator acting on the local doubled space at site $j$.
	The quantities $A_{\alpha \beta}(t)$ $(\alpha, \beta = I, X)$ can be written as
	\begin{equation}
		A_{II}(t) = \left( \mathrm{tr}_{4\times 4}[\Phi_j(t) \: S_j] \right)^L,
	\end{equation}
	\begin{equation}
		A_{XI}(t) = \left( \mathrm{tr}_{4\times 4}[\Phi_j(t) \: S_j \: (\sigma_j^x \otimes I_j)] \right)^L,
	\end{equation}
	\begin{equation}
		A_{IX}(t) = \left( \mathrm{tr}_{4\times 4}[\Phi_j(t) \: S_j \: (I_j \otimes \sigma_j^x)] \right)^L,
	\end{equation}
	\begin{equation}
		A_{XX}(t) = \left( \mathrm{tr}_{4\times 4}[\Phi_j(t) \: S_j \: (\sigma_j^x \otimes \sigma_j^x)] \right)^L,
	\end{equation}
	where $\mathrm{tr}_{4\times 4}$ represents the trace over the local doubled space (a four-dimensional space).
	
	By substituting the ansatz \eqref{single_site_mean_field_ansatz} into Eq.~\eqref{master_equation_swap_operator} and tracing out all sites except $j$, we obtain the effective local master equation for $\Phi_j(t)$:
	\begin{equation}
		\frac{d\Phi_j(t)}{dt} = i \sum_{\alpha=1,2} [H_{j, \text{eff}}^{(\alpha)}, \Phi_j(t)] + \gamma \sum_{\alpha=1,2} \left( \sigma_{j, \alpha}^x \Phi_j(t) \sigma_{j, \alpha}^x - \Phi_j(t) \right).
	\end{equation}
	Here, the inter-site interaction is replaced by a mean-field Hamiltonian, 
	\begin{equation}
		H_{j, \text{eff}}^{(\alpha)} = -J \sigma_{j, \alpha}^z \langle \sigma_{j-1, \alpha}^z \rangle - J \sigma_{j, \alpha}^z \langle \sigma_{j+1, \alpha}^z \rangle - h \sigma_{j, \alpha}^x,
	\end{equation}
	where the expectation value representing the effective field is given by $\langle \sigma_{k, \alpha}^z \rangle = \text{tr}_{4\times 4}[\sigma_{k, \alpha}^z \Phi_k(t)]/\text{tr}_{4\times 4}[\Phi_k(t)]$.
	
	Crucially, the initial condition $\Phi_j(0) = S_j$ is invariant under the local $\mathbb{Z}_2$ parity operation $\sigma_{j}^x \otimes \sigma_{j}^x$ acting on both copies.
	This symmetry ensures that the local magnetization $\langle \sigma_{k, \alpha}^z \rangle$ remains zero at all times during the dynamics if we assume that the system exhibits no conventional spontaneous symmetry breaking.
	Consequently, the inter-site interaction drops out of the mean-field Hamiltonian, yielding the explicit solution:
	\begin{equation}
		\Phi(t) = \frac{1}{2} \left[ I \otimes I + \sigma^x \otimes \sigma^x + e^{-4\gamma t} (\sigma^y \otimes \sigma^y + \sigma^z \otimes \sigma^z) \right],
	\end{equation}
	where the site index is omitted.
	Note that $\Phi(t)$ is independent of the transverse field $h$.
	This leads to
	\begin{equation}
		A_{II}(t) = [2(1 + e^{-4\gamma t})]^L,
	\end{equation}
	\begin{equation}
		A_{XX}(t) = [2(1 - e^{-4\gamma t})]^L,
	\end{equation}
	\begin{equation}
		A_{XI}(t) = A_{IX}(t) = 0.
	\end{equation}
	Thus we end up with
	\begin{equation}
		A(t) = 2^{L-2} \left[ (1 + e^{-4\gamma t})^L + (1 - e^{-4\gamma t})^L \right].
	\end{equation}
	Since $A(t)$ monotonically decreases from $A(0)=2^{2(L-1)}$ to $A(\infty)=2^{L-1}$, $A(t)^{1/(L-1)}/2$ never crosses the critical threshold $1$, implying $t_c=\infty$.
	Thus, the single-site mean-field approximation fails to capture the dynamical phase transition.
	The underlying physical reason is that the absence of inter-site interactions prevents the system from reaching the true infinite-temperature steady state in the mean-field dynamics.

	\subsection{Two-site cluster mean-field approximation}
	
	To overcome the limitation of the single-site mean-field approximation, we employ a cluster mean-field approximation.
	We partition the one-dimensional chain of even length $L$ into $L/2$ disjoint two-site clusters.
	The global operator $\mathcal{S}(t)$ is then approximated as a tensor product of the cluster operators:
	\begin{equation}
		\mathcal{S}(t) \simeq \bigotimes_{k=1}^{L/2} \Psi_{2k-1, 2k}(t),
	\end{equation}
	where $\Psi_{j, j+1}(t)$ acts on the doubled Hilbert space of the adjacent sites $j$ and $j+1$ (a 16-dimensional space).
	Substituting this ansatz into $A_{\alpha \beta}(t)$ $(\alpha, \beta = I, X)$, we obtain
	\begin{equation}
		A_{II}(t) = \left( \text{tr}_{16\times 16} [\Psi_{j, j+1}(t) \: S_{j} \: S_{j+1}] \right)^{L/2},
	\end{equation}
	\begin{equation}
		A_{XI}(t) = \left( \text{tr}_{16\times 16} [\Psi_{j, j+1}(t) \: S_{j} \: S_{j+1} \: (\sigma_j^x \otimes I_j) \: (\sigma_{j+1}^x \otimes I_{j+1})] \right)^{L/2},
	\end{equation}
	\begin{equation}
		A_{IX}(t) = \left( \text{tr}_{16\times 16} [\Psi_{j, j+1}(t) \: S_{j} \: S_{j+1} \: (I_j \otimes \sigma_j^x) \: (I_{j+1} \otimes \sigma_{j+1}^x)] \right)^{L/2},
	\end{equation}
	\begin{equation}
		A_{XX}(t) = \left( \text{tr}_{16\times 16} [\Psi_{j, j+1}(t) \: S_{j} \: S_{j+1} \: (\sigma_j^x \otimes \sigma_j^x) \: (\sigma_{j+1}^x \otimes \sigma_{j+1}^x)] \right)^{L/2},
	\end{equation}
	where $\text{tr}_{16\times 16}$ denotes the trace over the two-site doubled space.
	
	The time evolution of the cluster operator $\Psi_{j, j+1}(t)$ is governed by the master equation projected onto the two-site doubled space.
	The effective generator for the cluster incorporates the exact intra-cluster interaction alongside the mean-field coupling between adjacent clusters.
	By the same symmetry argument as in the single-site case, the inter-cluster mean fields vanish.
	Thus, the evolution equation simplifies to
	\begin{equation}
		\frac{d}{dt}\Psi_{j, j+1}(t) = i \sum_{\alpha=1,2} [H_{j,j+1}^{(\alpha)}, \Psi_{j, j+1}(t)] + \gamma \sum_{\alpha=1,2} \sum_{k=j, j+1} \left( \sigma_{k, \alpha}^x \Psi_{j, j+1}(t) \sigma_{k, \alpha}^x - \Psi_{j, j+1}(t) \right),
		\label{two_site_master_equation}
	\end{equation}
	where
	\begin{equation}
		H_{j,j+1}^{(\alpha)} = -J \sigma_{j, \alpha}^z \sigma_{j+1, \alpha}^z - h (\sigma_{j, \alpha}^x + \sigma_{j+1, \alpha}^x)
	\end{equation}
	is the two-site Hamiltonian acting on the $\alpha$-th copy of the doubled space.
	
	\begin{figure}
		\centering
		\includegraphics[width=0.6\textwidth]{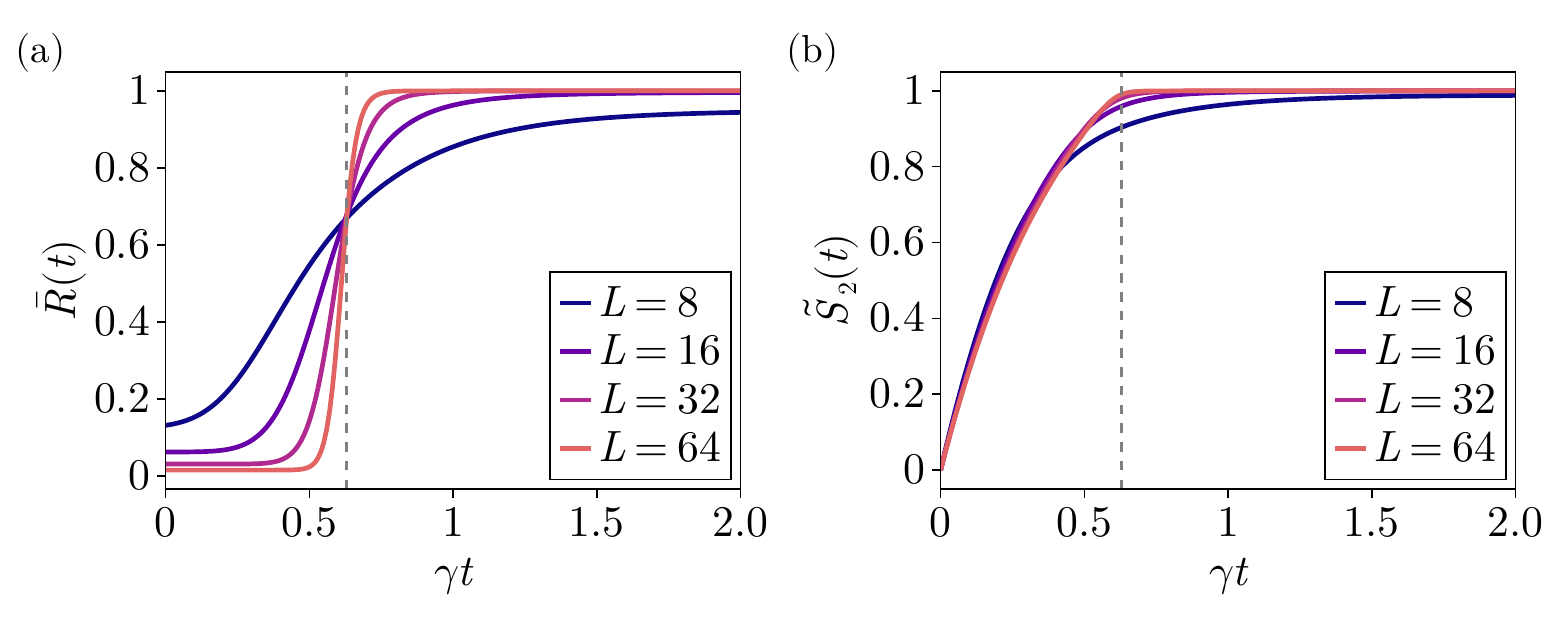}
		\caption{Dynamical phase transition captured by the two-site cluster mean-field approximation.
			(a) Time evolution of the R\'enyi-2 correlator $\bar{R}(t)$ for different system sizes $L=8, 16, 32$, and $64$.
			The parameters are set to $J/\gamma=1$ and $h=0$.
			A discontinuous jump emerges at the critical time $t_c$, signaling the onset of SWSSB.
			The vertical dashed line represents the crossing point, estimated as $\gamma t_c \simeq 0.63$.
			(b) Time evolution of the normalized R\'enyi-2 entropy $\tilde{S}_2(t) = S_2(t)/[(L-1) \ln 2]$. 
			The entropy exhibits a linear-like growth followed by a sharp saturation (kink) at the same critical time $t_c$.}
		\label{fig:renyi_entropy_mean_field}
	\end{figure}
	
	By numerically integrating Eq.~\eqref{two_site_master_equation} with the initial condition $\Psi_{j, j+1}(0) = S_j \: S_{j+1}$, one can directly evaluate $A(t)$.
	Figure \ref{fig:renyi_entropy_mean_field}(a) shows the spatially averaged R\'enyi-2 correlator $\bar{R}(t)$ calculated from Eq.~\eqref{spatially_averaged_renyi_correlator}.
	As the system size increases, a discontinuous jump clearly emerges at the critical time $\gamma t_c \simeq 0.63$, which is close to that obtained from the exact integration of the full master equation, $\gamma t_c \simeq 0.58$.
	Furthermore, by using Eq.~(7) of the main text, the R\'enyi-2 entropy is given by
	\begin{equation}
		S_2(t)=-\ln \mathrm{Tr}[\rho(t)^2] = -\ln \frac{1 + A(t)/2^{L-1}}{2^{L-1} + 1}.
	\end{equation}
	Figure \ref{fig:renyi_entropy_mean_field}(b) shows the normalized R\'enyi-2 entropy $\tilde{S}_2(t) = S_2(t)/[(L-1) \ln 2]$.
	The entropy exhibits an almost linear growth followed by saturation with a kink at the critical time $t_c$ same as that of the corresponding correlator.
	These behaviors are fully consistent with the exact numerical results presented in the main text.

	\section{Variance of purity}
	\label{sec:variance}
	
	In the main text, we have used the following approximations for the ensemble averages:
	\begin{equation}
		\mathbb{E}_{\psi_0}\left[ \frac{\mathrm{Tr}[\rho(t) \sigma_{i}^z \sigma_{j}^z \rho(t) \sigma_{i}^z \sigma_{j}^z]}{\mathrm{Tr}[\rho(t)^2]} \right] \simeq \frac{\mathbb{E}_{\psi_0}[\mathrm{Tr}[\rho(t) \sigma_{i}^z \sigma_{j}^z \rho(t) \sigma_{i}^z \sigma_{j}^z]]}{\mathbb{E}_{\psi_0}[\mathrm{Tr}[\rho(t)^2]]},
	\end{equation}
	\begin{equation}
		\mathbb{E}_{\psi_0}[\ln \mathrm{Tr}[\rho(t)^2]] \simeq \ln \mathbb{E}_{\psi_0}[\mathrm{Tr}[\rho(t)^2]],
	\end{equation}
	where $\mathbb{E}_{\psi_0}[\cdots]$ denotes the ensemble average over the random initial states.
	For these approximations to be valid, the sample-to-sample fluctuations of the relevant quantities, $\mathrm{Tr}[\rho(t)^2]$ and $\mathrm{Tr}[\rho(t) \sigma_{i}^z \sigma_{j}^z \rho(t) \sigma_{i}^z \sigma_{j}^z]$, must be negligibly small compared to their expectation values.
	Here, we focus on the purity $\mathcal{P}(t) := \mathrm{Tr}[\rho(t)^2]$, and evaluate its variance:
	\begin{equation}
		\mathrm{Var}[\mathcal{P}(t)] := \mathbb{E}_{\psi_0}[\mathcal{P}(t)^2] - \mathbb{E}_{\psi_0}[\mathcal{P}(t)]^2.
	\end{equation}
	In this section, we show that the relative variance of the purity, $\mathrm{Var}[\mathcal{P}(t)]/\mathbb{E}_{\psi_0}[\mathcal{P}(t)]^2$, is exponentially suppressed with respect to the system size $L$.
	This firmly establishes the self-averaging property underlying our analytical framework.
	
	To ensure analytical tractability, we restrict our analysis to the purely dephasing dynamics, neglecting inter-spin interactions and the transverse field.
	The corresponding master equation simplifies to
	\begin{equation}
		\frac{d\rho}{dt} = \gamma \sum_{j=1}^{L} \left( \sigma_{j}^{x} \rho \sigma_{j}^{x} - \rho \right).
		\label{master_equation_purely_dephasing}
	\end{equation}
	We assume that the initial state $\rho_0 = |\psi_0\rangle\langle\psi_0|$ is a pure state sampled uniformly at random from the entire Hilbert space (i.e., a Haar-random state).
	Using the swap operator $\mathcal{S}$, the purity $\mathcal{P}(t)$ can be written as
	\begin{equation}
		\mathcal{P}(t) = \mathrm{Tr}_{12} [(\rho(t) \otimes \rho(t)) \: \mathcal{S}] = \mathrm{Tr}_{12} [(\rho_0 \otimes \rho_0) \: \mathcal{S}(t)],
	\end{equation}
	where the time-evolved swap operator $\mathcal{S}(t)$ is defined by Eq.~\eqref{time_evolved_swap_operator}.
	Thus, the first and second moments of the purity are given by
	\begin{equation}
		\mathbb{E}_{\psi_0}[\mathcal{P}(t)] = \mathrm{Tr}_{12} [\mathbb{E}_{\psi_0}[\rho_0^{\otimes 2}] \: \mathcal{S}(t)],
	\end{equation}
	\begin{equation}
		\mathbb{E}_{\psi_0}[\mathcal{P}(t)^2] = \mathrm{Tr}_{1234} [\mathbb{E}_{\psi_0}[\rho_0^{\otimes 4}] \: (\mathcal{S}(t) \otimes \mathcal{S}(t))],
	\end{equation}
	where $\mathrm{Tr}_{1234}$ represents the trace over the four-copy Hilbert space $\mathcal{H}^{\otimes 4}$.
	
	The $k$th moment of a Haar-random pure state is given by
	\begin{equation}
		\mathbb{E}_{\psi_0}[\rho_0^{\otimes k}] = \frac{P_k}{D_k}, \qquad D_k = \binom{D+k-1}{k},
	\end{equation}
	where $P_k$ is the projection operator onto the completely symmetric subspace of $\mathcal{H}^{\otimes k}$ and $D=2^L$ is the dimension of $\mathcal{H}$ \cite{Collins-10, Hamma-12, Roberts-17, Nahum-18}.
	For example, for $k=2$, the projection operator is given by
	\begin{equation}
		P_2 = \frac{1}{2} (I \otimes I + \mathcal{S}),
	\end{equation}
	yielding
	\begin{equation}
		\mathbb{E}_{\psi_0}[\rho_0^{\otimes 2}] = \frac{I \otimes I + \mathcal{S}}{D (D + 1)}.
	\end{equation}
	
	The ensemble average of $\mathcal{P}(t)$ is then given by
	\begin{equation}
		\mu(t) := \mathbb{E}_{\psi_0}[\mathcal{P}(t)] = \frac{\mathrm{Tr}_{12} [P_2 \: \mathcal{S}(t)]}{D_2}.
		\label{mu_averaged_purity}
	\end{equation}
	To evaluate the variance, it is convenient to introduce the operator
	\begin{equation}
		\Delta \mathcal{S}(t) := P_2 \: \mathcal{S}(t) \: P_2 - \mu(t) P_2,
	\end{equation}
	which satisfies $\mathrm{Tr}_{12} [\Delta \mathcal{S}(t)] = 0$ because $\mathrm{Tr}_{12} [P_2] = D_2$.
	By using the identity $P_4 \: (P_2 \otimes P_2) = P_4$, the second moment of the purity can be expanded as
	\begin{align}
		\mathbb{E}_{\psi_0}[\mathcal{P}(t)^2] &=  \frac{1}{D_4} \mathrm{Tr}_{1234} [P_4 \: (\mathcal{S}(t) \otimes \mathcal{S}(t))] \nonumber \\
		&=  \frac{1}{D_4} \mathrm{Tr}_{1234} [P_4 \: [(P_2 \: \mathcal{S}(t) \: P_2) \otimes (P_2 \: \mathcal{S}(t) \: P_2)]] \nonumber \\
		&= \frac{1}{D_4}  \mathrm{Tr}_{1234} [P_4 \: [ \Delta \mathcal{S}(t) \otimes \Delta \mathcal{S}(t) + \mu(t) (\Delta \mathcal{S}(t) \otimes P_2 + P_2 \otimes \Delta \mathcal{S}(t)) + \mu(t)^2 P_2 \otimes P_2]].
	\end{align}
	Here, note that
	\begin{equation}
		\mathrm{Tr}_{1234} [P_4 \: (P_2 \otimes P_2)] = D_4,
	\end{equation}
	\begin{equation}
		\mathrm{Tr}_{1234} [P_4 \: (\Delta \mathcal{S}(t) \otimes P_2)] = \mathrm{Tr}_{12} [\mathrm{Tr}_{34}[P_4] \: \Delta \mathcal{S}(t)] \propto \mathrm{Tr}_{12} [P_2 \: \Delta \mathcal{S}(t)] = 0.
	\end{equation}
	Thus, we have
	\begin{equation}
		\mathrm{Var}[\mathcal{P}(t)] = \frac{\mathrm{Tr}_{1234} [P_4 \: (\Delta \mathcal{S}(t) \otimes \Delta \mathcal{S}(t))]}{D_4}.
		\label{purity_variance_1}
	\end{equation}
	
	Before evaluating the variance in Eq.~\eqref{purity_variance_1}, we derive the explicit expression for $\mathcal{S}(t)$.
	The global swap operator is written as
	\begin{equation}
		\mathcal{S} = \sum_{\mathbf{s}, \mathbf{s}'} (\ket{\mathbf{s}} \otimes \ket{\mathbf{s}'}) (\bra{\mathbf{s}'} \otimes \bra{\mathbf{s}}),
	\end{equation}
	where $\mathbf{s} = (s_1, \ldots , s_L)$ denotes a spin configuration in the $x$-basis, with $s_i = \pm 1$ such that $\sigma_i^x \ket{\mathbf{s}} = s_i \ket{\mathbf{s}}$.
	Under the purely dephasing master equation \eqref{master_equation_purely_dephasing}, the time evolution map $\Lambda_t^\dag$ exponentially suppresses the off-diagonal components:
	\begin{equation}
		\Lambda_t^\dag[|\mathbf{s}\rangle \langle \mathbf{s}'|] = e^{-2\gamma t |\mathbf{s}-\mathbf{s}'|} |\mathbf{s}\rangle \langle \mathbf{s}'|,
	\end{equation}
	where $|\mathbf{s}-\mathbf{s}'|$ represents the Hamming distance (i.e., the number of differing spins) between the spin configurations $\mathbf{s}$ and $\mathbf{s}'$.
	Thus, the time-evolved swap operator becomes
	\begin{equation}
		\mathcal{S}(t) = \sum_{\mathbf{s}, \mathbf{s}'} e^{-4\gamma t |\mathbf{s}-\mathbf{s}'|} (\ket{\mathbf{s}} \otimes \ket{\mathbf{s}'}) (\bra{\mathbf{s}'} \otimes \bra{\mathbf{s}}).
		\label{time_evolved_swap_operator_purely_dephasing}
	\end{equation}
	Using this expression, we can exactly evaluate the trace:
	\begin{equation}
		\mathrm{Tr}_{12} [\mathcal{S} \: \mathcal{S}(t)] = \sum_{\mathbf{s}, \mathbf{s}'} e^{-4\gamma t |\mathbf{s}-\mathbf{s}'|} = D \sum_{d=0}^L \binom{L}{d} e^{-4\gamma t d} = D (1 + e^{-4\gamma t})^L.
	\end{equation}
	Thus, the averaged purity given by Eq.~\eqref{mu_averaged_purity} is calculated as
	\begin{equation}
		\mu(t) = \frac{\mathrm{Tr}_{12} [(I \otimes I + \mathcal{S}) \: \mathcal{S}(t)]}{D(D+1)} = \frac{1 + (1 + e^{-4\gamma t})^L}{D + 1}.
		\label{mu_averaged_purity_expression}
	\end{equation}
	
	We show that the partial trace of $\Delta \mathcal{S}(t)$ over the second Hilbert space vanishes: $\mathrm{Tr}_2[\Delta \mathcal{S}(t)] = 0$.
	Since $\mathcal{S}(t)$ and $\mathcal{S}$ commute, we have $P_2 \: \mathcal{S}(t) \: P_2 = \frac{1}{2}(\mathcal{S}(t) + \mathcal{S} \: \mathcal{S}(t))$.
	Thus, Eq.~\eqref{time_evolved_swap_operator_purely_dephasing} yields
	\begin{equation}
		P_2 \: \mathcal{S}(t) \: P_2 = \frac{1}{2} \sum_{\mathbf{s}, \mathbf{s}'} e^{-4\gamma t |\mathbf{s}-\mathbf{s}'|} [(\ket{\mathbf{s}} \otimes \ket{\mathbf{s}'}) (\bra{\mathbf{s}} \otimes \bra{\mathbf{s}'}) + (\ket{\mathbf{s}} \otimes \ket{\mathbf{s}'}) (\bra{\mathbf{s}'} \otimes \bra{\mathbf{s}})].
	\end{equation}
	Taking the partial trace over the second Hilbert space, we obtain
	\begin{equation}
		\mathrm{Tr}_2[P_2 \: \mathcal{S}(t) \: P_2] = \frac{1}{2} \sum_{\mathbf{s}, \mathbf{s}'} (e^{-4\gamma t |\mathbf{s}-\mathbf{s}'|} + \delta_{\mathbf{s}, \mathbf{s}'}) |\mathbf{s}\rangle \langle \mathbf{s}| = \frac{1}{2} [1 + (1 + e^{-4\gamma t})^L] I.
	\end{equation}
	On the other hand, the partial trace of the mean purity term gives
	\begin{equation}
		\mathrm{Tr}_2[\mu(t) P_2] = \frac{1}{2} \mu(t) (D+1) I.
	\end{equation}
	Comparing these two results with Eq.~\eqref{mu_averaged_purity_expression}, we have $\mathrm{Tr}_2[\Delta \mathcal{S}(t)] = 0$.
	
	Having established $\mathrm{Tr}_2[\Delta \mathcal{S}(t)] = 0$, we can drastically simplify the four-copy trace in Eq.~\eqref{purity_variance_1}.
	The symmetric projection operator $P_4$ is defined as the sum over all 24 permutations in the symmetric group $\mathcal{S}_4$, i.e., $P_4 = \frac{1}{24} \sum_{\pi \in \mathcal{S}_4} W_\pi$, where $W_\pi$ is the corresponding permutation operator on $\mathcal{H}^{\otimes 4}$.
	Since any permutation that involves a partial trace over a single copy vanishes, out of the 24 permutations, 20 terms vanish identically.
	The only surviving contributions arise from the four permutations that completely swap the pairs $\{1,2\}$ and $\{3,4\}$ (e.g., $W_{(13)(24)}$).
	Each of these four terms yields $\mathrm{Tr}_{12}[\Delta \mathcal{S}(t)^2]$.
	Thus, the variance in Eq.~\eqref{purity_variance_1} simplifies to
	\begin{equation}
		\mathrm{Var}[\mathcal{P}(t)] = \frac{4}{24 D_4} \mathrm{Tr}_{12} [\Delta \mathcal{S}(t)^2] = \frac{1}{6 D_4} \left( \mathrm{Tr}_{12} [P_2 \: \mathcal{S}(t)^2] - \mu(t)^2 D_2 \right).
	\end{equation}
	To evaluate $\mathrm{Tr}_{12} [P_2 \: \mathcal{S}(t)^2]$, we note
	\begin{equation}
		\mathcal{S}(t)^2 = \sum_{\mathbf{s}, \mathbf{s}'} e^{-8\gamma t |\mathbf{s}-\mathbf{s}'|} (\ket{\mathbf{s}} \otimes \ket{\mathbf{s}'}) (\bra{\mathbf{s}} \otimes \bra{\mathbf{s}'}).
	\end{equation}
	Taking the trace with $P_2 = \frac{1}{2}(I \otimes I + \mathcal{S})$, we evaluate the two resulting terms separately:
	\begin{equation}
		\mathrm{Tr}_{12} [\mathcal{S}(t)^2] = \sum_{\mathbf{s}, \mathbf{s}'} e^{-8\gamma t |\mathbf{s}-\mathbf{s}'|} = D(1 + e^{-8\gamma t})^L, \qquad \mathrm{Tr}_{12} [\mathcal{S} \: \mathcal{S}(t)^2] = D.
	\end{equation}
	Summing up these contributions, we obtain
	\begin{equation}
		\mathrm{Tr}_{12} [P_2 \: \mathcal{S}(t)^2] = \frac{D}{2} [1 + (1 + e^{-8\gamma t})^L].
	\end{equation}
	Finally, we arrive at the analytical expression for the variance of the purity:
	\begin{equation}
		\mathrm{Var}[\mathcal{P}(t)] = \frac{2}{(D+1)(D+2)(D+3)} \left[ 1 + (1 + e^{-8\gamma t})^L - \frac{[1 + (1 + e^{-4\gamma t})^L]^2}{D+1} \right].
		\label{purity_variance_2}
	\end{equation}
	From this expression, the relative variance scales asymptotically as
	\begin{equation}
		\frac{\mathrm{Var}[\mathcal{P}(t)]}{\mu(t)^2} \sim \frac{1}{D} \left[\frac{1 + e^{-8\gamma t}}{(1 + e^{-4\gamma t})^2}\right]^L.
	\end{equation}
	Since $(1 + e^{-8\gamma t}) / (1 + e^{-4\gamma t})^2 < 1$ for all $t \ge 0$, the relative purity fluctuations are exponentially suppressed with increasing system size $L$.
	
	\begin{figure}
		\centering
		\includegraphics[width=0.35\textwidth]{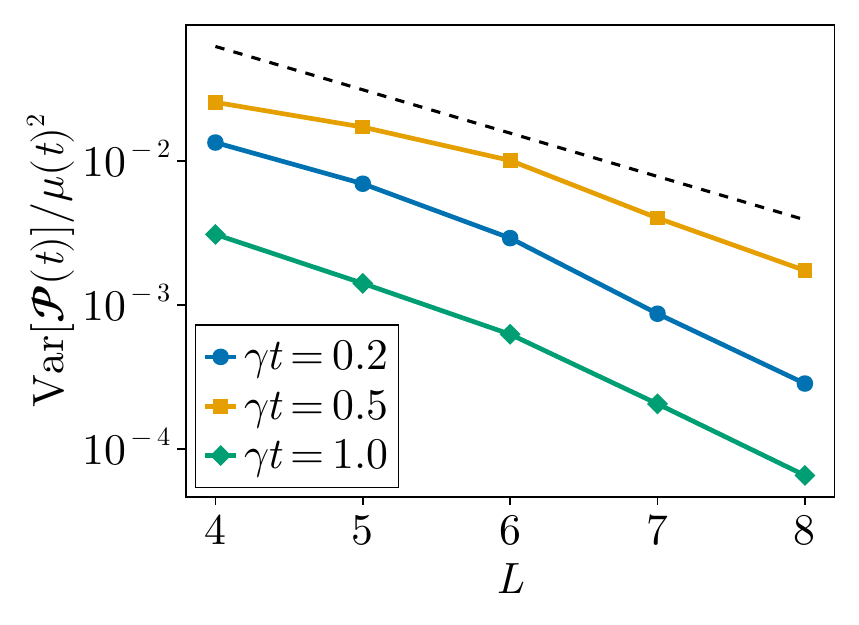}
		\caption{Relative variance of the purity, $\mathrm{Var}[\mathcal{P}(t)] / \mu(t)^2$, as a function of system size $L$ at various times $\gamma t = 0.2$, $0.5$, and $1.0$.
			The parameters are set to $J/\gamma=1$ and $h=0$.
			The black dashed line indicates $1/D = 2^{-L}$ for a guide to the eye.
			The data confirm that the purity fluctuations remain exponentially suppressed with increasing $L$ even in the presence of inter-spin interactions.}
		\label{fig:purity_variance}
	\end{figure}
	
	Note that the purely dephasing model analyzed above does not exhibit a dynamical transition to the SWSSB phase.
	To verify that the purity fluctuations remain negligible even in the presence of inter-spin interactions, we numerically evaluate the relative variance of the purity, $\mathrm{Var}[\mathcal{P}(t)] / \mu(t)^2$, for the same setup described in the main text.
	Figure \ref{fig:purity_variance} shows this relative variance as a function of system size $L$ at representative times.
	Recall that the critical time for this transition is $\gamma t_c \approx 0.5$.
	Although the magnitude of the fluctuations increases near the transition point, the data demonstrate that they remain exponentially suppressed with increasing $L$. 
	This numerical evidence confirms that the self-averaging property is robustly maintained across the entire dynamical evolution, including the critical regime.

	\section{Details of Numerical Simulations}
	\label{sec:num_details}
	
	In this section, we provide the technical details of the numerical simulations for the dephasing Ising model.
	
	\subsection{Setup}
	
	We consider a one-dimensional transverse-field Ising model of length $L$ with periodic boundary conditions. 
	The Hamiltonian is given by
	\begin{equation}
		H = -J \sum_{j=1}^{L} \sigma_{j}^{z} \sigma_{j+1}^{z} - h \sum_{j=1}^{L} \sigma_{j}^{x}, \quad (\sigma_{L+1}^{\mu} = \sigma_{1}^{\mu}),
		\label{appendix_Hamiltonian_Ising}
	\end{equation}
	where $\sigma_{j}^{\mu}$ $(\mu=x,y,z)$ are the Pauli matrices at site $j$, and $J$ and $h$ denote the interaction and the transverse field strength, respectively. 
	The open quantum dynamics subject to dephasing noise along the $x$-direction is governed by the Lindblad master equation:
	\begin{equation}
		\frac{d\rho}{dt} = -i[H, \rho] + \gamma \sum_{j=1}^{L} \left( \sigma_{j}^{x} \rho \sigma_{j}^{x} - \rho \right).
		\label{appendix_master_equation_Ising}
	\end{equation}
	This dissipative evolution preserves the strong $\mathbb{Z}_2$ symmetry associated with the global parity operator $X = \prod_{j=1}^L \sigma_j^x$.
	The time evolution of the density matrix $\rho(t)$ is computed via direct numerical integration of the Lindblad master equation using the standard Runge-Kutta method.
	Since the transverse field $h$ does not qualitatively alter the nature of SWSSB transitions, we set $J/\gamma = 1$ and $h/\gamma = 0$.
	
	The initial state is prepared as a strongly symmetric Haar-random pure state. 
	We first generate a random superposition state $|\tilde{\psi}_0\rangle = \sum_{\boldsymbol{s}} c(\boldsymbol{s})|\boldsymbol{s}\rangle$ in the computational $z$-basis, where $\boldsymbol{s}$ is a spin configuration of length $L$ and the coefficients $c(\boldsymbol{s})$ are independently sampled from a complex Gaussian distribution with zero mean and unit variance.
	To enforce the strong $\mathbb{Z}_2$ symmetry, we explicitly project the state into the even-parity sector as $|\psi_0\rangle \propto |\tilde{\psi}_0\rangle + X|\tilde{\psi}_0\rangle$ and normalize it to obtain $\rho_0 = |\psi_0\rangle\langle\psi_0|$.
	
	To detect the SWSSB order, we compute the spatially averaged fidelity correlator $\bar{F}$ and the R\'enyi-2 correlator $\bar{R}$. Using the local order parameter $\sigma_i^z$, these quantities are defined as
	\begin{align}
		\bar{F}(t) &= \frac{1}{L^2} \sum_{i,j=1}^{L} \mathrm{Tr}\sqrt{\sqrt{\rho(t)} \sigma_{i}^{z} \sigma_{j}^{z} \rho(t) \sigma_{j}^{z} \sigma_{i}^{z} \sqrt{\rho(t)}}, \\
		\bar{R}(t) &= \frac{1}{L^2} \sum_{i,j=1}^{L} \frac{\mathrm{Tr}[\rho(t) \sigma_{i}^{z} \sigma_{j}^{z} \rho(t) \sigma_{j}^{z} \sigma_{i}^{z}]}{\mathrm{Tr}[\rho(t)^2]}.
	\end{align}
	In the long-time limit, the system relaxes to the infinite-temperature state $\rho(\infty)=(I + X) / 2^L$ within the symmetry sector, yielding $\bar{F}_\infty = \bar{R}_\infty = 1$.
	These quantities are averaged over multiple realizations of the initial state; the number of random samples is set to $N_{\text{samp}}=200$ for $L=6$, $N_{\text{samp}}=100$ for $L=7, 8$, $N_{\text{samp}}=50$ for $L=9, 10$, $N_{\text{samp}}=25$ for $L=11$, and $N_{\text{samp}}=10$ for $L=12$.

	\subsection{Estimation of critical times and data collapse}
	
	To determine the critical time $t_c$, we perform a systematic analysis of the crossing points. 
	For a given correlator (either $\bar{F}$ or $\bar{R}$), we identify the intersection time $t_{c}(L_i, L_j)$ for pairs of different system sizes $(L_i, L_j)$.
	The final critical time $t_c$ is estimated as the average of these pairwise crossing times. 
	
	Subsequently, we perform a finite-size scaling analysis using the ansatz:
	\begin{equation}
		\bar{F}(t, L) = \Phi((t - t_c)L^\alpha),
	\end{equation}
	where $\Phi(\cdot)$ is a universal scaling function. 
	To find the optimal scaling exponent $\alpha$, we employ a data collapse optimization procedure. 
	Specifically, we minimize the variance between the scaled curves restricted to the time window within the SWSSB phase ($t > t_c$).
	The optimal data collapse is achieved with $\alpha \approx 1$.
	This exponent implies that the transition width shrinks as $1/L$, indicating a discontinuous, step-function-like jump in the thermodynamic limit $L \to \infty$.
	
	\subsection{Preparation of thermal and MBL initial states}
	
	In the main text, we investigate the dynamics starting from a thermal state and a many-body localized (MBL) state.
	An MBL state represents a highly excited eigenstate that violates the eigenstate thermalization hypothesis and possesses area-law entanglement \cite{Nandkishore-15, Abanin-19}, in contrast to the volume-law entanglement of thermal states \cite{Deutsch-91, Srednicki-94, Rigol-08}.
	
	To systematically generate both the thermal and MBL initial states, we introduce an auxiliary non-integrable Hamiltonian with random magnetic fields:
	\begin{equation}
		H_{\text{test}} = - J \sum_{j=1}^L \sigma_j^z \sigma_{j+1}^z - \sum_{j=1}^L h_j^x \sigma_j^x - \sum_{j=1}^L h_j^z \sigma_j^z.
	\end{equation}
	The nature of the eigenstates of $H_{\text{test}}$ is controlled by the distribution of the random fields $h_j^x$ and $h_j^z$. 
	For the thermal initial state, we draw $h_j^x / J$ and $h_j^z / J$ from a Gaussian distribution with a mean of $1.0$ and a standard deviation of $0.1$.
	This weak disorder places the system deep in the chaotic regime, characterized by Wigner-Dyson level statistics. 
	We then select an eigenstate $|\tilde{\psi}_0\rangle$ from the middle 25\% of the energy spectrum.
	Conversely, for the MBL initial state, we sample the random fields $h_j^x / J$ and $h_j^z / J$ from a Gaussian distribution with zero mean and a standard deviation of $10.0$.
	This strong disorder drives the system into the localized phase, characterized by Poisson level statistics.
	Similarly, we select an eigenstate $|\tilde{\psi}_0\rangle$ from the middle 25\% of the spectrum.
	
	In both cases, to ensure that the initial state strictly satisfies the strong symmetry condition, we explicitly project the chosen eigenstate into the even-parity sector using the global parity operator $X$: $|\psi_0\rangle \propto |\tilde{\psi}_0\rangle + X|\tilde{\psi}_0\rangle$.
	The subsequent time evolution of the density matrix $\rho_0 = |\psi_0\rangle\langle\psi_0|$ is governed by the disorder-free Lindblad master equation defined by Eqs.~\eqref{appendix_Hamiltonian_Ising} and \eqref{appendix_master_equation_Ising}.
	The physical observables are averaged over multiple disorder realizations of $H_{\text{test}}$; the number of random samples is set to $N_{\text{samp}}=1000$ for $L=6, 7$, $N_{\text{samp}}=500$ for $L=8, 9$, $N_{\text{samp}}=200$ for $L=10$, $N_{\text{samp}}=100$ for $L=11$, and $N_{\text{samp}}=50$ for $L=12$.

	\section{Results for the hard-core boson model}
	\label{sec:hard_core_boson}
	
	To corroborate the universality of our result, we consider a one-dimensional hard-core boson model with strong $U(1)$ symmetry.
	Each lattice site can be occupied by at most one boson.
	The Hamiltonian $H$ describes the hopping of bosons on a one-dimensional lattice of length $L$ with periodic boundary conditions: 
	\begin{equation}
		H = -J \sum_{j=1}^L (b_j^\dagger b_{j+1} + b_{j+1}^\dagger b_j), \quad (b_{L+1}=b_1),
	\end{equation}
	where $b_j$ is the annihilation operator of a boson at site $j$ and $J$ is the hopping amplitude.
	The open quantum dynamics subject to dephasing noise is governed by the Lindblad master equation:
	\begin{equation}
		\frac{d\rho}{dt} = -i[H, \rho] + \gamma \sum_{j=1}^{L} \left( n_{j} \rho n_{j} - \frac{1}{2} \{n_j, \rho \} \right),
	\end{equation}
	where $n_j = b_j^\dagger b_j$ is the local number operator. 
	This form of dissipation, which corresponds to the continuous measurement of local particle density, is experimentally realizable in ultracold atomic gases using, e.g., off-resonant light scattering \cite{Pichler-10, Pichler-13, Sarkar-14, Luschen-17, Bouganne-20}.
	Since the Hamiltonian and the Lindblad operators commute with the total particle number operator $N = \sum_{j=1}^L n_j$, the dynamics preserve the strong $U(1)$ symmetry.
	
	The numerical simulation is performed on the Hilbert space with a fixed particle number $N$, which has a dimension $D = \binom{L}{N}$. 
	The initial strongly $U(1)$-symmetric Haar-random pure state is generated by sampling a random superposition state $|\psi_0\rangle = \sum_{\boldsymbol{n}} c(\boldsymbol{n})|\boldsymbol{n}\rangle$, where $\boldsymbol{n} = (n_1, \dots, n_L)$ denotes the occupation numbers satisfying $\sum_i n_i = N$. 
	The complex coefficients $c(\boldsymbol{n})$ are drawn from a Gaussian distribution, followed by normalization to obtain $\rho_0 = |\psi_0\rangle\langle\psi_0|$.
	
	\begin{figure}
		\centering
		\includegraphics[width=0.6\textwidth]{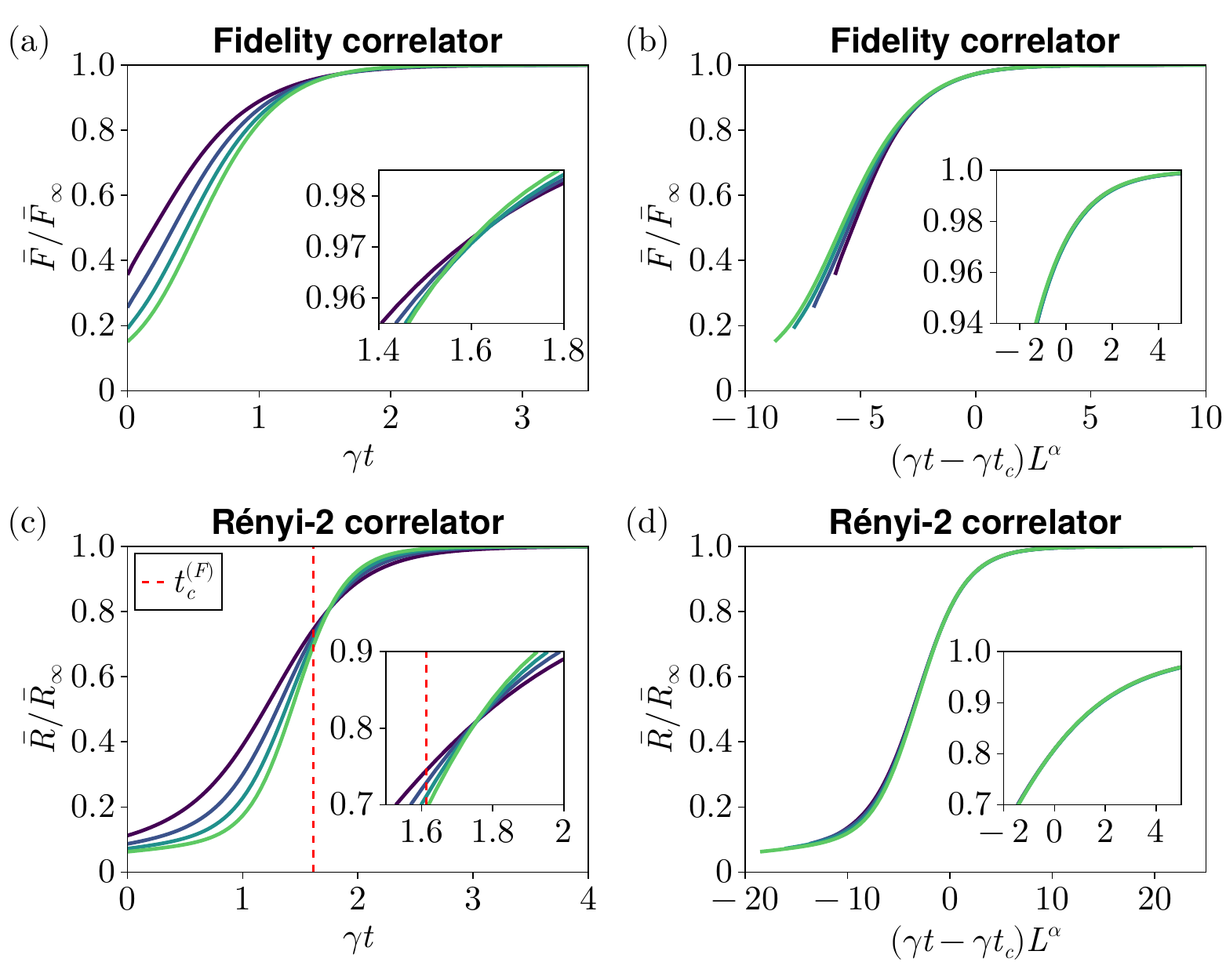}
		\caption{Dynamical SWSSB transitions in the hard-core boson model.
			(a) Time evolution of the normalized fidelity correlator $\bar{F}/\bar{F}_\infty$.
			The hopping amplitude is set to $J/\gamma=1$.
			Colors from dark to light represent system sizes $L = 8, 10, 12,$ and $14$ at half-filling ($N/L = 0.5$).
			The inset highlights the crossing point of the trajectories for different $L$, identifying the critical time $t_c^{(F)}$ of the dynamical transition.
			(b) Finite-size scaling analysis of the data in (a). 
			All curves collapse onto a single universal scaling function when plotted against the rescaled time $(\gamma t - \gamma t_c)L^\alpha$ with $\alpha =0.64$.
			(c) Time evolution of the normalized R\'enyi-2 correlator $\bar{R}/\bar{R}_\infty$.
			The vertical red dashed line indicates the critical time $t_c^{(F)}$ extracted from the fidelity correlator.
			The inset highlights the crossing point of $\bar{R}/\bar{R}_\infty$, identifying the critical time $t_c^{(R)}$, which occurs later than $t_c^{(F)}$.
			(d) Finite-size scaling analysis of the data in (c) using a scaling exponent $\alpha = 0.89$.}
		\label{fig:HCB_fidelity_renyi}
	\end{figure}
	
	For the hard-core bosons, the local order parameter is the annihilation operator $b_j$. The spatially averaged fidelity and R\'enyi-2 correlators are defined as
	\begin{align}
		\bar{F}(t) &= \frac{1}{L^2} \sum_{i,j=1}^{L} \mathrm{Tr}\sqrt{\sqrt{\rho(t)} b_{i} b_{j}^{\dagger} \rho(t) b_{j} b_{i}^{\dagger} \sqrt{\rho(t)}}, \\
		\bar{R}(t) &= \frac{1}{L^2} \sum_{i,j=1}^{L} \frac{\mathrm{Tr}[\rho(t) b_{i} b_{j}^{\dagger} \rho(t) b_{j} b_{i}^{\dagger}]}{\mathrm{Tr}[\rho(t)^2]}.
	\end{align}
	In the steady state, the system reaches the infinite-temperature state $\rho_\infty = I/D$ within the fixed-$N$ sector. 
	The infinite-time limits of the correlators are analytically calculated as
	\begin{equation}
		\bar{F}_\infty = \bar{R}_\infty = \frac{N + N(L-N)}{L^2}.
	\end{equation}
	Notably, unlike the Ising model, the steady-state values $\bar{F}_\infty$ and $\bar{R}_\infty$ depend on the system size $L$ for a fixed filling fraction $N/L$.
	To ensure a consistent comparison across different sizes, we focus on the normalized correlators, $\bar{F}(t)/\bar{F}_\infty$ and $\bar{R}(t)/\bar{R}_\infty$.
	
	Figure \ref{fig:HCB_fidelity_renyi} shows the time evolution of the normalized fidelity correlator and the R\'enyi-2 correlator for system sizes $L=8, 10, 12, 14$ at a fixed filling $N/L = 0.5$.
	In Figs.~\ref{fig:HCB_fidelity_renyi}(a) and (c), the curves for different system sizes exhibit crossing points, from which the critical times are estimated as $\gamma t_c^{(F)} = 1.61$ for the fidelity correlator and $\gamma t_c^{(R)} = 1.75$ for the R\'enyi-2 correlator.
	In Figs.~\ref{fig:HCB_fidelity_renyi}(b) and (d), we perform a finite-size scaling analysis.
	When plotted against the rescaled time $(\gamma t - \gamma t_c)L^\alpha$, the curves for different system sizes collapse onto a single universal scaling function.
	Notably, the transition detected by the R\'enyi-2 correlator occurs later than that of the fidelity correlator, $t_c^{(F)} < t_c^{(R)}$.
	These numerical results are consistent with the behavior observed in the dephasing Ising model, supporting the universality of the discontinuous SWSSB transition.

	\section{Absence of the dynamical transition for non-thermal initial states}
	\label{sec:nonthermal}
	
	In the main text, we have demonstrated that a strongly symmetric random pure state exhibiting volume-law entanglement undergoes a discontinuous dynamical phase transition into the SWSSB phase. 
	To highlight the fundamental importance of the initial entanglement structure, in this section, we investigate the dynamics starting from non-thermal initial states that violate the ``volume-law and random'' condition. 
	We show that the sharp dynamical transition is absent for such initial states.
	In this section, we focus on the dynamics of the R\'enyi-2 correlator in the dephasing Ising model.
	
	\subsection{GHZ state}
	
	As a representative example of a strongly symmetric state with area-law entanglement, we consider the Greenberger-Horne-Zeilinger (GHZ) state defined in the $x$-basis:
	\begin{equation}
		|\psi_{\text{GHZ}}\rangle = \frac{1}{\sqrt{2}} \left( |+x\rangle^{\otimes L} + |-x\rangle^{\otimes L} \right),
	\end{equation}
	where $|\pm x\rangle = (|\uparrow\rangle \pm |\downarrow\rangle) / \sqrt{2}$. 
	For an even number of sites $L$, this state belongs to the $+1$ symmetry sector of the global parity operator $X = \prod_{j=1}^L \sigma_j^x$, satisfying the strong symmetry requirement $X|\psi_{\text{GHZ}}\rangle = |\psi_{\text{GHZ}}\rangle$.
	The R\'enyi-2 correlator for the GHZ state is given by
	\begin{equation}
		R_O^{(2)}(i, j) = 
		\begin{cases}
			1, & (i = j), \\
			0, & (i \neq j),
		\end{cases}
	\end{equation}
	which leads to a spatially averaged correlator of $\bar{R}(t=0) = 1/ L$.
	Thus, the GHZ state does not exhibit SWSSB.
	
	\begin{figure}
		\centering
		\includegraphics[width=0.6\textwidth]{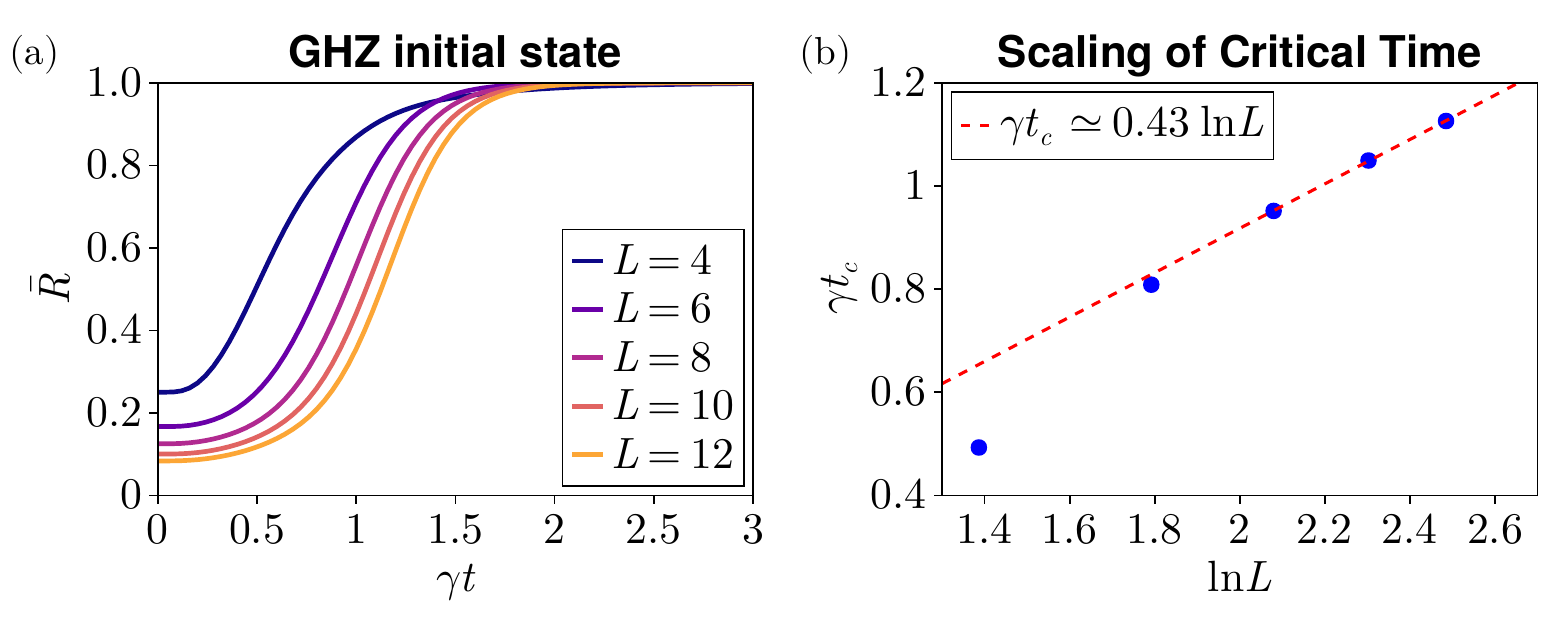}
		\caption{Absence of the dynamical transition for the GHZ initial state.
			(a) Time evolution of the spatially averaged R\'enyi-2 correlator $\bar{R}(t)$ for various system sizes $L=4, 6, 8, 10, 12$. 
			Unlike the case of random pure states, the correlators do not exhibit a system-size-independent crossing point. 
			(b) Finite-size scaling of the threshold time $t_c$, defined by the condition $\bar{R}(t_c) = 0.5$. 
			The dashed red line represents a linear fit, indicating a logarithmic divergence $t_c \propto \ln L$.}
		\label{fig:GHZ_initial_state}
	\end{figure}
	
	The time evolution of the spatially averaged R\'enyi-2 correlator $\bar{R}(t)$ starting from the GHZ state is shown in Fig.~\ref{fig:GHZ_initial_state}(a).
	In contrast to the Haar-random or thermal initial states discussed in the main text, the trajectories of $\bar{R}(t)$ for different system sizes $L$ do not intersect at a size-independent crossing point.
	While $\bar{R}(t)$ eventually approaches the steady-state value of unity, the time scale required for this growth depends on $L$.
	To quantify this size dependence, we define a threshold time $t_c$ at which the correlator reaches a value of $0.5$.
	As demonstrated in Fig.~\ref{fig:GHZ_initial_state}(b), $t_c$ increases monotonically with the system size.
	A linear fit to the data reveals a logarithmic divergence, 
	\begin{equation}
		t_c \propto \ln L, 
	\end{equation}
	implying that the time required to develop long-range correlations extends to infinity in the thermodynamic limit.
	In particular, in the thermodynamic limit, $\bar{R}(t)$ remains pinned at zero for any finite time.
	It should be noted that the logarithmic divergence of the critical time in one-dimensional systems has been reported in Ref.~\cite{Shu-26}.

	\subsection{Dicke state}
	
	As a next example of a non-thermal initial state, we consider the half-filled Dicke state in the $z$-basis.
	For a system of $L$ spins, the Dicke state $|L, k\rangle$ with $k$ excitations is defined as the joint eigenstate of the total spin operators $\bm{S}^2$ and $S^z$ (where $\bm{S} = \frac{1}{2}\sum_j \bm{\sigma}_j$) with eigenvalues $S(S+1) = \frac{L}{2}(\frac{L}{2}+1)$ and $M = k - L/2$, respectively.
	This state can be constructed by applying the collective spin raising operator $S^+ = \sum_{j=1}^L (\sigma_j^x + i\sigma_j^y)/2$ to the fully polarized state $\ket{\downarrow}^{\otimes L}$:
	\begin{equation}
		|\psi_{\text{Dicke}}\rangle \propto (S^+)^k \ket{\downarrow}^{\otimes L}.
	\end{equation}
	In the computational basis, this definition is equivalent to an equal-weight superposition of all basis states containing exactly $k$ up-spins:
	\begin{equation}
		|\psi_{\text{Dicke}}\rangle = \binom{L}{k}^{-1/2} \sum_{\{s\} | \sum_i s_i = k} |s_1, s_2, \dots, s_L\rangle,
	\end{equation}
	where $s_i \in \{0, 1\}$ corresponds to the local spin states $\ket{\downarrow}$ and $\ket{\uparrow}$, respectively.
	In particular, we focus on the half-filled case ($k=L/2$) for an even number of sites $L$.
	
	\begin{figure}
		\centering
		\includegraphics[width=0.6\textwidth]{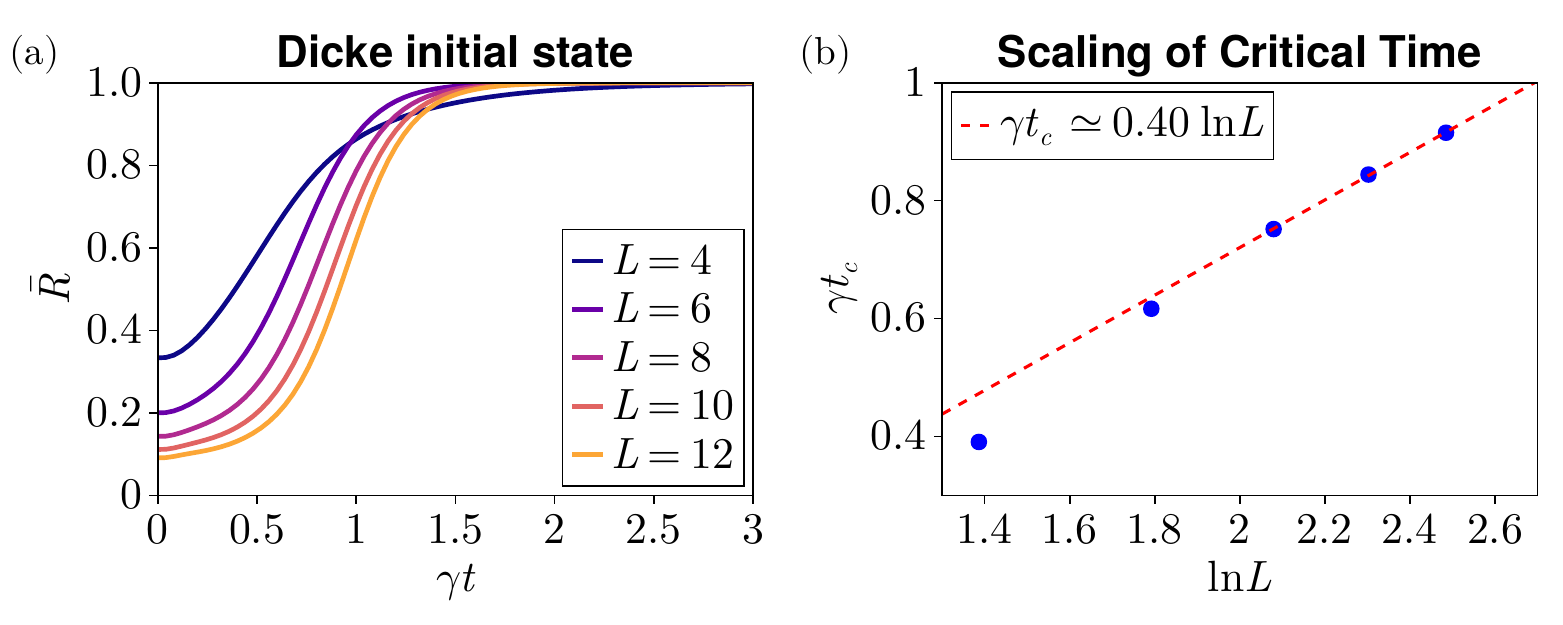}
		\caption{Absence of the dynamical transition for the Dicke initial state.
			(a) Time evolution of the spatially averaged R\'enyi-2 correlator $\bar{R}(t)$ for various system sizes $L=4, 6, 8, 10, 12$. 
			Unlike the case of random pure states, the correlators do not exhibit a system-size-independent crossing point. 
			(b) Finite-size scaling of the threshold time $t_c$, defined by the condition $\bar{R}(t_c) = 0.5$. 
			The dashed red line represents a linear fit, indicating a logarithmic divergence $t_c \propto \ln L$.}
		\label{fig:Dicke_initial_state}
	\end{figure}
	
	The half-filled Dicke state naturally satisfies the strong symmetry condition, $X|\psi_{\text{Dicke}}\rangle = |\psi_{\text{Dicke}}\rangle$.
	Furthermore, this state does not exhibit SWSSB.
	This can be understood from the strict conservation of the total magnetization. 
	Let $M_z = \sum_{j=1}^L \sigma_j^z$ be the total magnetization operator. 
	For the half-filled Dicke state, $M_z |\psi_{\text{Dicke}}\rangle = 0$, which implies
	\begin{equation}
		\langle \psi_{\text{Dicke}}| (M_z)^2 |\psi_{\text{Dicke}}\rangle = L + \sum_{i \neq j} \langle \sigma_i^z \sigma_j^z \rangle = 0.
	\end{equation}
	Thus, for all pairs $i \neq j$, we obtain $\langle \sigma_i^z \sigma_j^z \rangle = -1/(L-1)$. 
	This yields the following expression for the R\'enyi-2 correlator
	\begin{equation}
		R_O^{(2)}(i, j) = 
		\begin{cases}
			1, & (i = j), \\
			\frac{1}{(L-1)^2}, & (i \neq j).
		\end{cases}
	\end{equation}
	In the thermodynamic limit $L \to \infty$, the off-diagonal correlator vanishes as $O(L^{-2})$, confirming that the system does not exhibit SWSSB initially.
	
	While the Dicke state possesses highly non-local correlations, its entanglement entropy scales only logarithmically with the subsystem size, $S \sim \ln L$ \cite{Popkov-05, Kunimi-25}.
	The time evolution of the spatially averaged R\'enyi-2 correlator $\bar{R}(t)$ starting from the Dicke state is shown in Fig.~\ref{fig:Dicke_initial_state}.
	We find that the dynamics are qualitatively similar to those observed for the GHZ initial state: the correlators for different system sizes do not exhibit a size-independent crossing point, and its threshold time $t_c$ diverges logarithmically as $L$ increases.
	These results support the conclusion that the discontinuous dynamical phase transition into the SWSSB phase is absent for subvolume-law entangled states in the thermodynamic limit $L \to \infty$.
	
	\subsection{Rainbow state}
	
	To definitively isolate the essential ingredients required for the discontinuous dynamical transition, we investigate the dynamics starting from the rainbow state \cite{Langlett-22, Wildeboer-22, Chiba-24, Mohapatra-25}.
	Unlike the GHZ, MBL, and Dicke states discussed above, the rainbow state exhibits volume-law entanglement entropy for a standard spatial bipartition.
	
	Consider a one-dimensional chain of even length $L$. 
	The rainbow state is constructed such that each site $i$ in the left half ($1 \le i \le L/2$) forms a maximally entangled Bell pair with its mirror-image site $\tilde{i} = L - i + 1$ in the right half. 
	In the computational $z$-basis, this is written as a product of distant Bell pairs:
	\begin{equation}
		|\psi_{\text{rainbow}}\rangle = \bigotimes_{i=1}^{L/2} \frac{1}{\sqrt{2}} \left( |\uparrow\rangle_i |\uparrow\rangle_{\tilde{i}} + |\downarrow\rangle_i |\downarrow\rangle_{\tilde{i}} \right).
	\end{equation}
	Equivalently, this state is the equal-weight superposition of all computational basis states whose spin configurations form palindromes.
	
	This palindromic structure naturally ensures the strong $\mathbb{Z}_2$ symmetry, $X|\psi_{\text{rainbow}}\rangle = |\psi_{\text{rainbow}}\rangle$.
	Furthermore, the initial state does not exhibit SWSSB. 
	The two-point correlation $\langle \sigma_i^z \sigma_j^z \rangle$ is nonzero (equal to $1$) only if $i=j$ or $j=\tilde{i}$, and zero otherwise. 
	Consequently, the spatially averaged R\'enyi-2 correlator scales as $\bar{R}(0) = 2/L$, which vanishes in the thermodynamic limit.
	While the entanglement entropy of the rainbow state scales linearly with respect to subsystem size for a typical spatial cut, it strictly vanishes for a fine-tuned bipartition that groups the mirror pairs $(i, \tilde{i})$ together.
	Specifically, the rainbow state lacks the randomness present in Haar-random or thermal states.
	
	\begin{figure}
		\centering
		\includegraphics[width=0.6\textwidth]{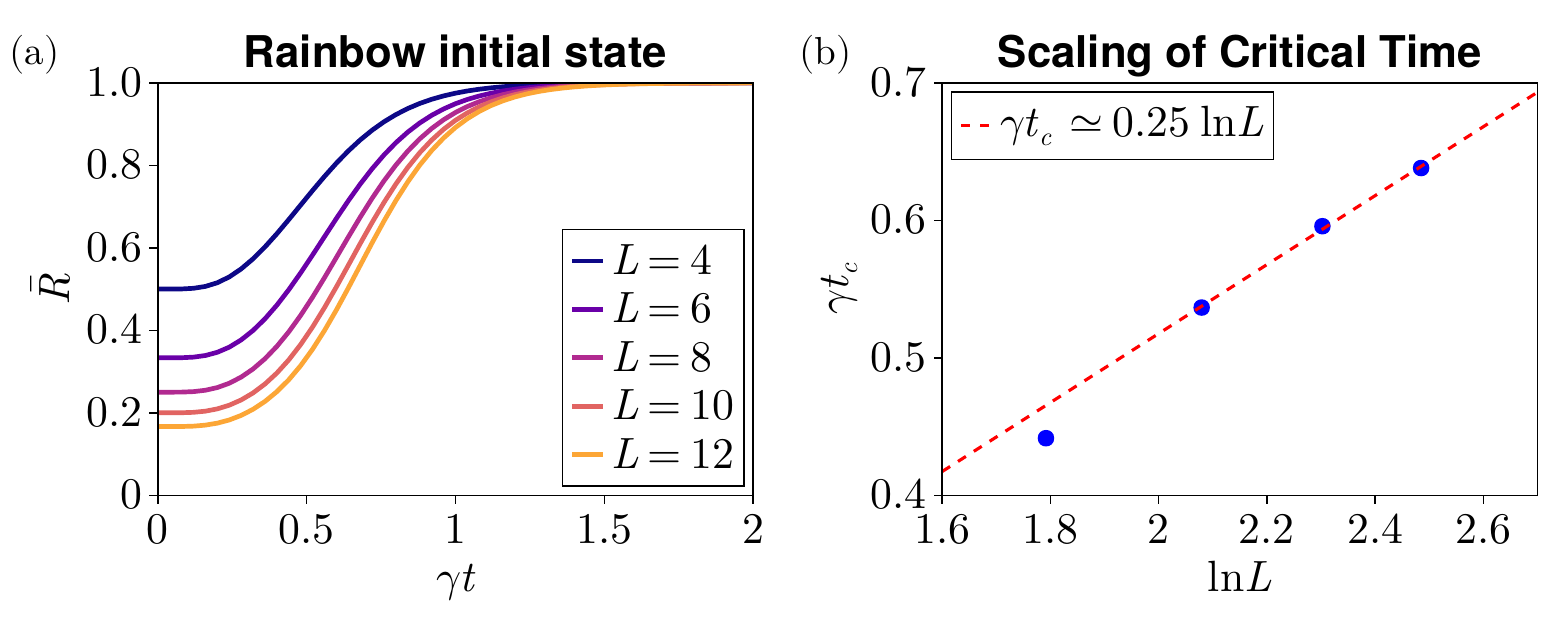}
		\caption{Absence of the dynamical transition for the rainbow initial state.
			(a) Time evolution of the spatially averaged R\'enyi-2 correlator $\bar{R}(t)$ for various system sizes $L=4, 6, 8, 10, 12$. 
			Unlike the case of random pure states, the correlators do not exhibit a system-size-independent crossing point. 
			(b) Finite-size scaling of the threshold time $t_c$, defined by the condition $\bar{R}(t_c) = 0.5$. 
			The dashed red line represents a linear fit, indicating a logarithmic divergence $t_c \propto \ln L$.}
		\label{fig:rainbow_initial_state}
	\end{figure}
	
	Figure \ref{fig:rainbow_initial_state} shows the time evolution of the spatially averaged R\'enyi-2 correlator $\bar{R}(t)$ starting from the rainbow state.
	Despite its volume-law entanglement, the time evolution of $\bar{R}(t)$ resembles that of the area-law GHZ state. 
	As seen in Fig.~\ref{fig:rainbow_initial_state}(a), the correlators for different system sizes fail to intersect at a size-independent crossing point. 
	Furthermore, the threshold time $t_c$ required for the correlator to reach $0.5$ diverges logarithmically with the system size, $t_c \propto \ln L$ [Fig.~\ref{fig:rainbow_initial_state}(b)].
	These results demonstrate that the presence of volume-law entanglement alone is insufficient to induce the discontinuous dynamical transition. 
	Instead, macroscopic randomness, such as that inherent in Haar-random or thermal pure states, constitutes an indispensable prerequisite for triggering the transition.
	

\begin{thebibliography}{99}
		\bibitem{Lee-23} Jong Yeon Lee, Chao-Ming Jian, and Cenke Xu, \textit{Quantum Criticality Under Decoherence or Weak Measurement,} PRX Quantum \textbf{4}, 030317 (2023).
		
		\bibitem{Sala-24} Pablo Sala, Sarang Gopalakrishnan, Masaki Oshikawa, and Yizhi You, \textit{Spontaneous strong symmetry breaking in open systems: Purification perspective,} Phys. Rev. B \textbf{110}, 155150 (2024).
		
		\bibitem{Lessa-25} Leonardo A. Lessa, Ruochen Ma, Jian-Hao Zhang, Zhen Bi, Meng Cheng, and Chong Wang, \textit{Strong-to-Weak Spontaneous Symmetry Breaking in Mixed Quantum States,} PRX Quantum \textbf{6}, 010344 (2025).
		
		\bibitem{ChongWang-26} Chong Wang, \textit{Strong-to-Weak Spontaneous Symmetry Breaking,} arXiv:2606.02555.
		
		\bibitem{Kuno-24} Yoshihito Kuno, Takahiro Orito, and Ikuo Ichinose, \textit{Strong-to-weak symmetry breaking states in stochastic dephasing stabilizer circuits,} Phys. Rev. B \textbf{110}, 094106 (2024).
		
		\bibitem{Guo-25} Yuxuan Guo, Sheng Yang, and Xue-Jia Yu, \textit{Quantum Strong-To-Weak Spontaneous Symmetry Breaking in Decohered One-Dimensional Critical States,} PRX Quantum \textbf{6}, 040311 (2025).
		
		\bibitem{Ellison-25} Tyler D. Ellison and Meng Cheng, \textit{Toward a Classification of Mixed-State Topological Orders in Two Dimensions,} PRX Quantum \textbf{6}, 010315 (2025).
		
		\bibitem{Luo-25} Ran Luo, Yi-Nan Wang, and Zhen Bi, \textit{Topological Holography for Mixed-State Phases and Phase Transitions,} PRX Quantum \textbf{6}, 040358 (2025).
		
		\bibitem{Ma-25} Ruochen Ma, Jian-Hao Zhang, Zhen Bi, Meng Cheng, and Chong Wang, \textit{Topological Phases with Average Symmetries: The Decohered, the Disordered, and the Intrinsic,} Phys. Rev. X \textbf{15}, 021062 (2025).
		
		\bibitem{Shah-25} Jeet Shah, Christopher Fechisin, Yu-Xin Wang, Joseph T. Iosue, James D. Watson, Yan-Qi Wang, Brayden Ware, Alexey V. Gorshkov, and Cheng-Ju Lin, \textit{Instability of steady-state mixed-state symmetry-protected topological order to strong-to-weak spontaneous symmetry breaking,} Quantum \textbf{9}, 1912 (2025).
		
		\bibitem{Orito-25} Takahiro Orito, Yoshihito Kuno, and Ikuo Ichinose, \textit{Strong and weak symmetries and their spontaneous symmetry breaking in mixed states emerging from the quantum Ising model under multiple decoherence,} Phys. Rev. B \textbf{111}, 054106 (2025).
		
		\bibitem{Zhang-25} Carolyn Zhang, Yichen Xu, Jian-Hao Zhang, Cenke Xu, Zhen Bi, and Zhu-Xi Luo, \textit{Strong-to-weak spontaneous breaking of 1-form symmetry and intrinsically mixed topological order,} Phys. Rev. B \textbf{111}, 115137 (2025).
		
		\bibitem{Huang-25} Xiaoyang Huang, Marvin Qi, Jian-Hao Zhang, and Andrew Lucas, \textit{Hydrodynamics as the effective field theory of strong-to-weak spontaneous symmetry breaking,} Phys. Rev. B \textbf{111}, 125147 (2025).
		
		\bibitem{Sa-25} Lucas S\'a and Benjamin B\'eri, \textit{Exactly solvable dissipative dynamics and one-form strong-to-weak spontaneous symmetry breaking in interacting two-dimensional spin systems,} Phys. Rev. B \textbf{112}, 144311 (2025).
		
		\bibitem{Gu-25} Ding Gu, Zijian Wang, and Zhong Wang, \textit{Spontaneous symmetry breaking in open quantum systems: Strong, weak, and strong-to-weak,} Phys. Rev. B \textbf{112}, 245123 (2025).
		
		\bibitem{Weinstein-25} Zack Weinstein, \textit{Efficient Detection of Strong-to-Weak Spontaneous Symmetry Breaking via the R\'enyi-1 Correlator,} Phys. Rev. Lett. \textbf{134}, 150405 (2025).
		
		\bibitem{Sun-25} Ning Sun, Pengfei Zhang, and Lei Feng, \textit{Scheme to Detect the Strong-to-Weak Symmetry Breaking via Randomized Measurements,} Phys. Rev. Lett. \textbf{135}, 090403 (2025).
		
		\bibitem{Feng-25} Xiaozhou Feng, Zihan Cheng, and Matteo Ippoliti, \textit{Hardness of Observing Strong-to-Weak Symmetry Breaking,} Phys. Rev. Lett. \textbf{135}, 200402 (2025).
		
		\bibitem{Liu-25} Zeyu Liu, Langxuan Chen, Yuke Zhang, Shuyan Zhou, and Pengfei Zhang, \textit{Diagnosing strong-to-weak symmetry breaking via Wightman correlators,} Commun. Phys. \textbf{8}, 274 (2025).
		
		\bibitem{Ando-26} Takamasa Ando, Shinsei Ryu, and Masataka Watanabe, \textit{Gauge theory and mixed state criticality,} Phys. Rev. B \textbf{113}, 115106 (2026).
		
		\bibitem{Hauser-26} Jacob Hauser, Kaixiang Su, Hyunsoo Ha, Jerome Lloyd, Thomas G. Kiely, Romain Vasseur, Sarang Gopalakrishnan, Cenke Xu, and Matthew P. A. Fisher, \textit{Strong-to-Weak Symmetry Breaking in Open Quantum Systems: From Discrete Particles to Continuum Hydrodynamics,} arXiv:2602.16045.
		
		\bibitem{Shu-26} Chang Shu, Kai Zhang, Zhu-Xi Luo, Yizhi You, and Kai Sun, \textit{Universal Dynamical Scaling of Strong-to-Weak Spontaneous Symmetry Breaking in Open Quantum Systems,} arXiv:2603.06363.
		
		\bibitem{Wang-26} Si Wang, Thomas G. Kiely, Dorothee Tell, Johannes Obermeyer, Marnix Barendregt, Petar Bojovi\'c, Philipp M. Preiss, Abhijat Sarma, Titus Franz, Matthew P. A. Fisher, Cenke Xu, and Immanuel Bloch, \textit{Observation of Strong-to-Weak Spontaneous Symmetry Breaking in a Dephased Fermi Gas,} arXiv:2604.16137.
		
		\bibitem{Teh-25} Hung-Hsuan Teh and Takahiro Orito, \textit{Krylov Complexity and Mixed-State Phase Transition,} arXiv:2510.22542.
		
		\bibitem{Ding-26} Yi-Ming Ding, Yuxuan Guo, Zhen Bi, and Zheng Yan, \textit{Strong-to-Weak Spontaneous Symmetry Breaking in a (2 + 1)D Transverse-Field Ising Model under Decoherence,} arXiv:2603.24342.
		
		\bibitem{Buca-12} Berislav Bu{\v{c}}a and Toma{\v{z}} Prosen, \textit{A note on symmetry reductions of the Lindblad equation: transport in constrained open spin chains,} New J. Phys. \textbf{14}, 073007 (2012).
		
		\bibitem{Albert-14} Victor V. Albert and Liang Jiang, \textit{Symmetries and conserved quantities in Lindblad master equations,} Phys. Rev. A \textbf{89}, 022118 (2014).
		
		\bibitem{Page-93} Don N. Page, \textit{Average Entropy of a Subsystem,} Phys. Rev. Lett. \textbf{71}, 1291 (1993).
		
		\bibitem{Breuer} Heinz-Peter Breuer and Francesco Petruccione, \textit{The Theory of Open Quantum Systems} (Oxford University Press, New York, 2002).
		
		\bibitem{Rivas} \'Angel Rivas and Susana F. Huelga, \textit{Open Quantum Systems} (Springer, Berlin, 2012).
		
		\bibitem{Sugiura-13} Sho Sugiura and Akira Shimizu, \textit{Canonical Thermal Pure Quantum State,} Phys. Rev. Lett. \textbf{111}, 010401 (2013).
		
		\bibitem{Mori-18} Takashi Mori, Tatsuhiko N Ikeda, Eriko Kaminishi, and Masahito Ueda, \textit{Thermalization and prethermalization in isolated quantum systems: a theoretical overview,} J. Phys. B: At. Mol. Opt. Phys. \textbf{51}, 112001 (2018).
		
		\bibitem{Divi-26} Francisco Divi, Leonardo A. Lessa, and Chong Wang, \textit{Local Strong-to-Weak Spontaneous Symmetry Breaking,} arXiv:2605.28967.
		
		\bibitem{Liu-26} Ruizhi Liu, Jinmin Yi, and Dominic V. Else, \textit{A local description of strong symmetries and strong-to-weak symmetry breaking in quantum many-body systems,} arXiv:2605.28925.
		
		\bibitem{Zhang-26} Carolyn Zhang, \textit{Local diagnostics for strong-to-weak spontaneous symmetry breaking and non-equilibrium phase transitions,} arXiv:2605.29113.
		
		\bibitem{Tang-26} Yicheng Tang, Pradip Kattel, and J. H. Pixley, \textit{A mean-field description of strong-to-weak symmetry breaking in the monitored three-dimensional Bose-Hubbard model,} arXiv:2606.02713.
		
		\bibitem{Kaufman-16} Adam M. Kaufman, M. Eric Tai, Alexander Lukin, Matthew Rispoli, Robert Schittko, Philipp M. Preiss, and Markus Greiner, \textit{Quantum thermalization through entanglement in an isolated many-body system,} Science \textbf{353}, 794 (2016).
		
		\bibitem{SM} See Supplemental Material for additional details, which includes a theoretical derivation of the transition using the cluster mean-field approximation, an analytical estimate of the fluctuations in purity, technical specifics of the simulations, numerical results for the hard-core boson model, and evidence for the absence of SWSSB in non-thermal initial states.
		
		\bibitem{Pichler-10} Hannes Pichler, Andrew J. Daley, and Peter Zoller, \textit{Nonequilibrium dynamics of bosonic atoms in optical lattices: Decoherence of many-body states due to spontaneous emission,} Phys. Rev. A \textbf{82}, 063605 (2010).
		
		\bibitem{Pichler-13} Hannes Pichler, Johannes Schachenmayer, Andrew J. Daley, and Peter Zoller, \textit{Heating dynamics of bosonic atoms in a noisy optical lattice,} Phys. Rev. A \textbf{87}, 033606 (2013).
		
		\bibitem{Sarkar-14} Saubhik Sarkar, Stephan Langer, Johannes Schachenmayer, and Andrew J. Daley, \textit{Light scattering and dissipative dynamics of many fermionic atoms in an optical lattice,} Phys. Rev. A \textbf{90}, 023618 (2014).
		
		\bibitem{Luschen-17} Henrik P. L\"uschen, Pranjal Bordia, Sean S. Hodgman, Michael Schreiber, Saubhik Sarkar, Andrew J. Daley, Mark H. Fischer, Ehud Altman, Immanuel Bloch, and Ulrich Schneider, \textit{Signatures of Many-Body Localization in a Controlled Open Quantum System,} Phys. Rev. X \textbf{7}, 011034 (2017).
		
		\bibitem{Bouganne-20} Rapha\"el Bouganne, Manel Bosch Aguilera, Alexis Ghermaoui, J\'er\^ome Beugnon, and Fabrice Gerbier, \textit{Anomalous decay of coherence in a dissipative many-body system}, Nat. Phys. \textbf{16}, 21 (2020).
		
		\bibitem{Bao-26} Ruicheng Bao, \textit{Initial-State Typicality in Quantum Relaxation,} Phys. Rev. Lett. \textbf{136}, 070402 (2026).
		
		\bibitem{Collins-10} Beno\^it Collins and Ion Nechita, \textit{Random Quantum Channels I: Graphical Calculus and the Bell State Phenomenon,} Commun. Math. Phys. \textbf{297}, 345 (2010).
		
		\bibitem{Hamma-12} Alioscia Hamma, Siddhartha Santra, and Paolo Zanardi, \textit{Ensembles of physical states and random quantum circuits on graphs,} Phys. Rev. A \textbf{86}, 052324 (2012).
		
		\bibitem{Roberts-17} Daniel A. Roberts and Beni Yoshida, \textit{Chaos and complexity by design,} J. High Energy Phys. \textbf{04}, 121 (2017).
		
		\bibitem{Nahum-18}  Adam Nahum, Sagar Vijay, and Jeongwan Haah, \textit{Operator Spreading in Random Unitary Circuits,} Phys. Rev. X \textbf{8}, 021014 (2018).
		
		\bibitem{Deutsch-91} J. M. Deutsch, \textit{Quantum statistical mechanics in a closed system,} Phys. Rev. A \textbf{43}, 2046 (1991).
		
		\bibitem{Srednicki-94} Mark Srednicki, \textit{Chaos and quantum thermalization,} Phys. Rev. E \textbf{50}, 888 (1994).
		
		\bibitem{Rigol-08} Marcos Rigol, Vanja Dunjko, and Maxim Olshanii, \textit{Thermalization and its mechanism for generic isolated quantum systems,} Nature \textbf{452}, 854 (2008).
		
		\bibitem{Nandkishore-15} Rahul Nandkishore and David A. Huse, \textit{Many-Body Localization and Thermalization in Quantum Statistical Mechanics,} Annu. Rev. Condens. Matter Phys. \textbf{6}, 15 (2015).
		
		\bibitem{Abanin-19} Dmitry A. Abanin, Ehud Altman, Immanuel Bloch, and Maksym Serbyn, \textit{Colloquium: Many-body localization, thermalization, and entanglement,} Rev. Mod. Phys. \textbf{91}, 021001 (2019).
		
		\bibitem{Popkov-05} Vladislav Popkov and Mario Salerno, \textit{Logarithmic divergence of the block entanglement entropy for the ferromagnetic Heisenberg model,} Phys. Rev. A \textbf{71}, 012301 (2005).
		
		\bibitem{Kunimi-25} Masaya Kunimi, Yusuke Kato, and Hosho Katsura, \textit{Systematic construction of asymptotic quantum many-body scar states and their relation to supersymmetric quantum mechanics,} Phys. Rev. Research \textbf{7}, 043107 (2025).
		
		\bibitem{Langlett-22} Christopher M. Langlett, Zhi-Cheng Yang, Julia Wildeboer, Alexey V. Gorshkov, Thomas Iadecola, and Shenglong Xu, \textit{Rainbow scars: From area to volume law}, Phys. Rev. B \textbf{105}, L060301 (2022).
		
		\bibitem{Wildeboer-22} Julia Wildeboer, Christopher M. Langlett, Zhi-Cheng Yang, Alexey V. Gorshkov, Thomas Iadecola, and Shenglong Xu, \textit{Quantum many-body scars from Einstein-Podolsky-Rosen states in bilayer systems,} Phys. Rev. B \textbf{106}, 205142 (2022).
		
		\bibitem{Chiba-24} Yuuya Chiba and Yasushi Yoneta, \textit{Exact Thermal Eigenstates of Nonintegrable Spin Chains at Infinite Temperature,} Phys. Rev. Lett. \textbf{133}, 170404 (2024).
		
		\bibitem{Mohapatra-25} Sashikanta Mohapatra, Sanjay Moudgalya, and Ajit C. Balram, \textit{Exact Volume-Law Entangled Zero-Energy Eigenstates in a Large Class of Spin Models,} Phys. Rev. Lett. \textbf{134}, 210403 (2025).
		
		\bibitem{Chen-25} Langxuan Chen, Ning Sun, and Pengfei Zhang, \textit{Strong-to-weak symmetry breaking and entanglement transitions}, Phys. Rev. B \textbf{111}, L060304 (2025).
		
		\bibitem{Chen-20} Yiming Chen, Xiao-Liang Qi, and Pengfei Zhang, \textit{Replica wormhole and information retrieval in the SYK model coupled to Majorana chains,} J. High Energy Phys. \textbf{06} (2020) 121.
		
		\bibitem{Wang-24} Hanteng Wang, Chang Liu, Pengfei Zhang, and Antonio M. Garc{\'\i}a-Garc{\'\i}a, \textit{Entanglement transition and replica wormholes in the dissipative Sachdev-Ye-Kitaev model,} Phys. Rev. D \textbf{109}, 046005 (2024).
		
		\bibitem{data} Taiki Haga, GitHub, 2026, https://github.com/taiki-haga/discontinuous-SWSSB-transition		
	\end{thebibliography}
\end{document}